\documentclass{sigchi}

\CopyrightYear{2020}
\setcopyright{acmlicensed}
\doi{https://doi.org/10.1145/3313831.3376839}
\isbn{978-1-4503-6708-0/20/04}
\conferenceinfo{CHI'20,}{April  25--30, 2020, Honolulu, HI, USA}
\acmPrice{\$15.00}

\newcommand{\todo}[1]{}
\renewcommand{\todo}[1]{{\color{red} \textbf{TODO: {#1}}}}

\newcommand{\fix}[1]{}
\renewcommand{\fix}[1]{{\color{orange} \textbf{FIXME: {#1}}}}

\newcommand{\cit}[1]{}
\renewcommand{\cit}[1]{{\color{green} \textbf{CITEME: {#1}}}}

\newcommand{\system}{Mix\&Match\xspace}
\newcommand{\systemEmph}{\emph{\system}\xspace}

\newcommand{\paradigm}{\emph{''pick-and-choose''}\xspace}
\newcommand{\mmf}{\emph{MyMiniFactory}\xspace}
\newcommand{\tv}{\emph{Thingiverse}\xspace}

\usepackage{balance}       %
\usepackage{graphicx}      %
\usepackage[T1]{fontenc}   %
\usepackage{txfonts}
\usepackage{mathptmx}
\usepackage[pdflang={en-US},pdftex]{hyperref}
\usepackage{xcolor}
\usepackage{booktabs}
\usepackage{textcomp}
\usepackage{mwe}
\usepackage{subcaption}
\usepackage{microtype}        %
\usepackage{ccicons}          %

\def\plaintitle{\system: Towards Omitting Modelling Through In-situ Remixing of Model Repository Artifacts in Mixed Reality}

\def\emptyauthor{}
\def\plainkeywords{Personal Fabrication; Model Repositories; Mixed Reality; In-Situ Modelling; In-Situ Previews; 3D Printing}

\makeatletter
\def\url@leostyle{%
  \@ifundefined{selectfont}{
    \def\UrlFont{\sf}
  }{
    \def\UrlFont{\small\bf\ttfamily}
  }}
\makeatother
\def\pprw{8.5in}
\def\pprh{11in}

\definecolor{linkColor}{RGB}{6,125,233}
\hypersetup{%
  pdftitle={\plaintitle},
  pdfauthor={\emptyauthor},
  pdfkeywords={\plainkeywords},
  pdfdisplaydoctitle=true, %
  bookmarksnumbered,
  pdfstartview={FitH},
  colorlinks,
  citecolor=black,
  filecolor=black,
  linkcolor=black,
  urlcolor=linkColor,
  breaklinks=true,
  hypertexnames=false
}

\usepackage{cuted}
\usepackage{capt-of}

\PassOptionsToPackage{hyphens}{url}
\toappear{\scriptsize Permission to make digital or hard copies of all or part of this work for personal or classroom use is granted without fee provided that copies are not made or distributed for profit or commercial advantage and that copies bear this notice and the full citation on the first page. Copyrights for components of this work owned by others than ACM must be honored. Abstracting with credit is permitted. To copy otherwise, or republish, to post on servers or to redistribute to lists, requires prior specific permission and/or a fee. Request permissions from permissions@acm.org. \\
{\emph{CHI '20, April 25--30, 2020, Honolulu, HI, USA.} } \\
Copyright is held by the owner/author(s). Publication rights licensed to ACM. \\
ACM ISBN 978-1-4503-6708-0/20/04\ ...\$15.00.\\
http://dx.doi.org/10.1145/3313831.3376839}

\begin{document}

\title{\plaintitle}

\numberofauthors{1}
\author{%
  \alignauthor{Evgeny Stemasov, Tobias Wagner, Jan Gugenheimer\footnotemark, Enrico Rukzio\\
    \affaddr{Institute of Media Informatics}\\
    \affaddr{Ulm University, Germany}\\
    \email{<\textit{fistname}>.<\textit{lastname}>@uni-ulm.de}}\\
}

\maketitle

\begin{strip}\centering
    \includegraphics[width=1\textwidth]{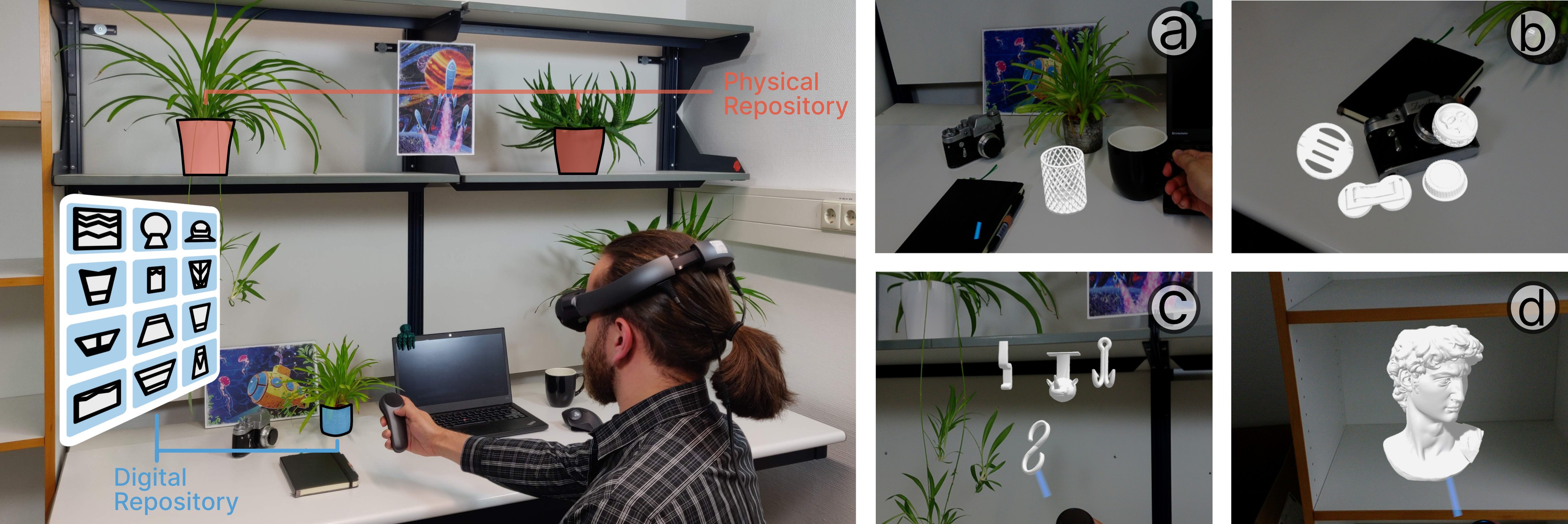}
    \captionof{figure}{
        \systemEmph is a proof-of-concept implementation of a tool for personal fabrication that leverages rich model repositories, while retaining artifact adaptivity with respect to the users' personal physical context. 
        It treats the users' environment as a repository of its own, and allows for in-situ previews and alterations.
        Left: Designs for a flower pot can be gathered from a repository like Thingiverse, or be derived from existing flower pots, if they have been acquired in sufficient detail.
        Right: Previewing the scale of future artifacts, relative to existing ones (a), comparing alternatives in-situ (b, c), and verifying clearance for future artifacts (d).
    \label{fig:teaser}}
\end{strip}

\begin{abstract}

The accessibility of tools to model artifacts is one of the core driving factors for the adoption of Personal Fabrication.
Subsequently, model repositories like Thingiverse became important tools in (novice) makers' processes.
They allow them to shorten or even omit the design process, offloading a majority of the effort to other parties.
However, steps like measurement of surrounding constraints (e.g., clearance) which exist only inside the users' environment, can not be similarly outsourced.
We propose \systemEmph, a mixed-reality-based system which allows users to browse model repositories, preview the models in-situ, and adapt them to their environment in a simple and immediate fashion. 
\systemEmph aims to provide users with CSG operations which can be based on both virtual and real geometry.
We present interaction patterns and scenarios for \systemEmph, arguing for the combination of mixed reality and model repositories. This enables almost modelling-free personal fabrication for both novices and expert makers.

\end{abstract}




\section{Introduction and Motivation}
\label{sec:introduction}

\renewcommand*{\thefootnote}{\fnsymbol{footnote}}
\addtocounter{footnote}{1}\footnotetext{now at T\'{e}l\'{e}com Paris/IP-Paris}
\renewcommand*{\thefootnote}{\arabic{footnote}}

Personal fabrication continues to spread across various usage contexts, ranging from low-volume prototyping, makerspaces, and users' homes.
The cost of required hardware (e.g., 3D printers or CNC-mills) is continuously decreasing~\cite{shewbridgeEverydayMakingIdentifying2014}, while the variety of available devices is ever-increasing.
This allows experienced users to create various artifacts specifically tailored to their needs.
While the artifacts' quality may not always match industry-grade production processes, they may still fulfill functional requirements. 
Personal fabrication therefore sees use beyond toys and trinkets, and instead enables practical changes in households~\cite{chenRepriseDesignTool2016,shewbridgeEverydayMakingIdentifying2014}.
Users may repair broken objects, and likewise create entirely new artifacts, for instance for home improvement ~\cite{chenRepriseDesignTool2016} or enhanced accessibility~\cite{buehlerSharingCaringAssistive2015,mcdonaldUncoveringChallengesOpportunities2016,interikeasystemsb.v.IkeaThisAbles2019}.
All such use-cases \emph{empower} users to alter their physical environments and democratize the process of fabrication.

However, the successful usage of such devices (e.g., 3D printers, CNC-mills) often requires some degree of knowledge of complex tools~\cite{baudischPersonalFabrication2017,leeUsabilityPrinciplesBest2010}.
CAD/CAM software was initially transferred from industrial usage, and only later experienced simplification aimed at novices~\cite{baudischPersonalFabrication2017}.
Alternatively, it is possible to replace most, if not all, modelling with the usage of a model repository, where other users make their designs freely available to the public~\cite{alcockBarriersUsingCustomizing2016,buehlerSharingCaringAssistive2015,roumenGrafterRemixing3DPrinted2018}.
Users then omit modelling, and instead browse the repository for artifacts (i.e., solutions) to fabricate.
While open model repositories provide users with ready-to-print artifacts, they exhibit conceptual limitations.
The design effort is offloaded to other parties, but the knowledge and understanding of problem specifics (e.g., clearances or proportions) remains with the respective users and their unique requirements.
More importantly, the entire physical context remains with the user and has to be mapped and measured~\cite{mahapatraBarriersEndUserDesigners2019}.
For objects that are not explicitly standardized or hard to gauge, this quickly becomes an issue for novices and potentially time-consuming for more experienced makers~\cite{kimUnderstandingUncertaintyMeasurement2017}.
Subsequently, these measurements and specified constraints can be mis-measured~\cite{kimUnderstandingUncertaintyMeasurement2017} or missed by the user, requiring additional iterations~\cite{teibrichPatchingPhysicalObjects2015}.
Our work aims to bridge this disconnect between the physical space of the end-user and the space of the model repository. 
We argue for an easy \emph{in-situ} \paradigm fabrication paradigm largely omitting the need for more complex modelling tools and operations.

We propose \system, a Mixed-Reality-based tool, which aims to leverage outsourced design effort through model repositories, while retaining relevant and easy in-situ adaptations.
\system was implemented using a Magic Leap ML1 augmented reality (AR) head-mounted display (HMD) and the \mmf repository. 
We provide a visual interface to the model repository, allowing users to search for artifacts and browse through results in place (Figure \ref{fig:teaser}b-c).
The models can be compared in-situ and altered with modifications like scaling to ensure both aesthetic and functional fit to the environment (Figure \ref{fig:teaser}a-d).
As an AR-headset already provides depth-sensing, it can incorporate the physical environment into the selection (Figure \ref{fig:teaser}) and alteration process (Figure \ref{fig:teaser}c).
To allow for simple adaptations to the environment of the user, \system provides Constructive Solid Geometry (CSG) operations, that can be based on digital artifacts, retrieved from the model repository.
These Boolean operations are also applicable to real artifacts, as found in the users' immediate vicinity, if they have been acquired appropriately.
This allows the user to subtract geometry of a shelf from another part to ensure a friction fit or make a digital copy of a physical artifact and thereby treating the physical environment as a repository of its own (Figure \ref{fig:teaser}).
All features are aimed at increased ease of use through outsourced effort and the omission of modelling to create an easier access for novices and accelerate the process for more experienced makers.

Instead of limiting users of a model repository to pre-defined approaches, \system encourages the practice of \emph{in-situ remixing} of artifacts, while treating the users' physical environment in a similar fashion to a digital one.
We propose and argue for an \emph{in-situ \paradigm}-based personal fabrication paradigm.
Model repositories like \tv or \mmf already provide readily available, free designs.
However, with \emph{in-situ \paradigm}, we want to compensate for some of the inherent disadvantages their approach of outsourced design implicates: a disconnect between the physical environment of the user and the repositories' functionality.
The contributions of this work are:
\begin{itemize}
    \item Proof-of-concept implementation of \system, a Mixed-Reality-based tool that allows users to preview and alter model repository artifacts in-situ.
    \item The notion of an \emph{in-situ} \paradigm-based personal fabrication paradigm and set of application scenarios and interaction flows for \system and comparable systems.
\end{itemize}

Ultimately, a system like \system allows to outsource many, if not all, parts of personal fabrication that do not have to be inherently personal.
Design/modelling effort is outsourced to a crowd of experienced makers.
Measurement is offloaded to a hardware system (e.g., depth cameras of an AR-headset).
Fabrication of the artifact itself \emph{can} be likewise offloaded to an external service.
With these components delegated to other, often more competent parties, novices and experienced makers alike may achieve fitting results with fewer interaction cycles.

\section{Related Work}
\system builds upon multiple directions of research: fabrication or design with mixed reality, personal fabrication for novices and personal fabrication that interacts with its physical counterparts, along with research concerned with the use and improvement of model repositories.
\label{sec:relatedwork}

\subsection{Fabrication in or with Mixed Reality}
    Mixed or augmented reality, along with all related technologies, has shown to be a promising tool for personal fabrication activities.
    As such, it is able to provide previews of models or enable easier in-situ modelling of artifacts.
    Milette and McGuffin presented DualCAD, which combined a smartphone device and an HMD for 3D-modelling~\cite{milletteDualCADIntegratingAugmented2016}.
    Mixed reality also allows users to interactively influence and guide a fabrication process, as for instance shown by
    Peng et al. with RoMA, where the authors combined augmented reality and a robotic arm \cite{pengRoMAInteractiveFabrication2018}.
    Yamaoka and Kakehi presented MiragePrinter, where an aerial imaging plate combined the fabricated result of a 3D printer with output from modelling software~\cite{yamaokaMiragePrinterInteractiveFabrication2016}.
    This allowed users to rely on physical artifacts as guides and interactively control the fabrication process~\cite{yamaokaMiragePrinterInteractiveFabrication2016}. 

    Mixed reality can also be used as a guidance for manual tasks done by the user.
    Yue et al. presented WireDraw, which supports users in the task of drawing in mid-air with a 3D-pen~\cite{yueWireDraw3DWire2017}.
    For subtractive manufacturing, Hattab and Taubin presented a method to support users carving an object with information projected onto the workpiece \cite{hattabRoughCarving3D2019}.
    ExoSkin by Gannon et al. aided users with projected toolpaths to fabricate intricate shapes on the body -- a complex but relevant feature for personal fabrication~\cite{gannonExoSkinOnBodyFabrication2016}.
    Jeong et al. applied this concept to the design process of linkage mechanisms in Mechanism Perfboard~\cite{jeongMechanismPerfboardAugmented2018}, while M\"{u}ller et al. aimed to improve the ease of use of CNC mills with augmented reality~\cite{mullerCaMeaCameraSupportedWorkpiece2018}.
    Weichel et al. presented  MixFab, which used mixed reality to provide users with a tool that actively includes scanned real-world artifacts and gesture-based modelling in the process~\cite{weichelMixFabMixedrealityEnvironment2014}.
    
    The aforementioned works have in common that they situate modelling work in a spatial context, ideally co-locating it with relevant real-world features and improving processes of measurement and understanding.
    \system differs from them primarily in two ways: 1) It is not meant to be confined to a static setup. 
    2) it is not meant to be a ''pure'' design tool that essentially makes users ''start from scratch''.
    Instead, \system relies on outsourced design effort, as provided by model repositories, to allow users to omit modelling as such.

\subsection{Fabrication for Novices}
    It is important to consider the aspect that tools used for personal fabrication did not start out as explicitly novice-friendly.
    Research therefore focused on accessibility of the modelling processes itself.
    Drill Sergeant aimed to equip novices with a set of tools that are able to provide feedback and generally support the fabrication process~\cite{schoopDrillSergeantSupporting2016}, while CopyCAD by Follmer et al. allowed users to copy features from arbitraty objects to reference in a CNC-milling setup~\cite{follmerCopyCADRemixingPhysical2010}.
    Makers' Marks by Savage et al. allowed users without technical knowledge to design functional artifacts through sculpting a shape and annotating it with the desired features~\cite{savageMakersMarksPhysical2015}.
    Turning coarse input into viable designs through sketches was also a prior topic.
    SketchChair, was a tool to let novices design and verify chairs~\cite{saulSketchChairAllinoneChair2011}.
    Lau et al. generalized this concept later, aiming at arbitrary objects to be personalised~\cite{lauSketchingPrototypingPersonalised2012}.
    Yung et al. presented Printy3D, which combined two paradigms to ease the process: the design happens in-situ and also employs tangibility in the interface~\cite{yungPrinty3DInsituTangible2018}.
    
    With \system, we similarly aim to simplify the process of personal fabrication, but without the goal to simplify \emph{modelling} tools.
    Instead, we aim to omit modelling (in its established sense) completely, while retaining relevant abilities to configure and alter artifacts.

\subsection{Model Repositories and Remixes}
    Prior research has also focused on the usage and extension of model repositories.
    Alcock et al. categorized issues that novices or other users may have when it comes to usage and adaptation of model repository artifacts, identifying missing information, customization and customizability as issues present on \tv~\cite{alcockBarriersUsingCustomizing2016}.
    Novices to 3D-printing and associated processes were the topic of Hudson et al., who identified common challenges like missing domain knowledge or the inability to customize existing designs \cite{hudsonUnderstandingNewcomers3D2016}.
    ''Parameterized Abstractions of Reusable Things'' were introduced as a framework by Hofmann et al. to counteract a disconnect between designed artifacts and their intended functionality~\cite{hofmannGreaterSumIts2018}.
    Kim et al. aimed to improve on the error-prone process of measuring artifacts to be references in 3D-printing by introducing adjustable inserts or replaceable parts~\cite{kimUnderstandingUncertaintyMeasurement2017}.
    
    The concept of remixing model repository artifacts is an important process in online 3D-printing communities~ \cite{oehlbergPatternsPhysicalDesign2015}.
    Roumen et al. presented Grafter, a tool to aid in the process of remixing machines~\cite{roumenGrafterRemixing3DPrinted2018}, while Follmer and et al. presented tools to do so for toys~\cite{follmerKidCADDigitallyRemixing2012} and other physical artifacts~\cite{follmerCopyCADRemixingPhysical2010}.
    Lindlbauer and Wilson, in contrast, presented Remixed Reality, where mediated reality served as a tool to alter one's own physical context~\cite{lindlbauerRemixedRealityManipulating2018} from and in a digital environment.
    
    \system aims to provide a novel, situated interface to model repositories, bridging the gap between outsourced designs and the users' physical context, allowing in-situ previewing and remixing.
    
\subsection{Fabrication for and with Real-world Artifacts}
    Personal fabrication may yield various artifacts: decorative figures, household items, replicas of existing objects, props, tools etc.
    No result is going to exist ''in a vacuum'' -- every artifact interacts with its environment.
    This concept was specified by Ashbrook et al. as augmented fabrication~\cite{ashbrookAugmentedFabricationCombining2016, mahapatraBarriersEndUserDesigners2019}, and was also prominent part of prior research \cite{follmerCopyCADRemixingPhysical2010,weichelMixFabMixedrealityEnvironment2014,savageMakersMarksPhysical2015}.
    Yamada et al. presented ReFabricator, a tool to actively integrate real-world objects as material in fabricated artifacts ~\cite{yamadaReFabricatorIntegratingEveryday2016}.
    In contrast to that, FusePrint by Zhu et al. incorporated real-world objects as references in a stereolithography printing process~\cite{zhuFusePrintDIY5D2016}, while Huo et al. leveraged real-world features as an input for 3D design with Window-Shaping~\cite{huoWindowShaping3DDesign2017}.
    Lau et al. relied on a photograph to create fitting objects~\cite{lauModelingincontextUserDesign2010}.
    Ramakers et al. presented RetroFab, which allowed users to retroactively alter and enhance physical interfaces like desk lamps or toasters~\cite{ramakersRetroFabDesignTool2016}.
    ThisAbles by Ikea presents 3D-printable improvements to furniture, to accommodate for users' special needs~\cite{interikeasystemsb.v.IkeaThisAbles2019}.
    Chen et al. presented a set of tools to combine real-world artifacts with 3D-printed ones. Reprise focused on customizeable adaptations for everyday tools and objects~\cite{chenRepriseDesignTool2016}, Encore dealt with attachments and their fabrication~\cite{chenEncore3DPrinted2015}, while Medley treated everyday objects as materials to augment 3D-printed objects~\cite{chenMedleyLibraryEmbeddables2018}.
    The previously mentioned MixFab by Weichel et al. likewise incorporates real-world artifacts as a support for operations~\cite{weichelMixFabMixedrealityEnvironment2014}.
    In contrast to these approaches, AutoConnect by Koyama et al. mostly automates the process of modelling 3D-printable connectors for various objects~\cite{koyamaAutoConnectComputationalDesign2015}.
    
    \system embraces the procedures presented here, extending them by leveraging model repository artifacts, while simultaneously providing ways to embed the physical context into the process by allowing for in-situ previews and alterations like CSG referencing the users' physical context.
\section{Concept and Interaction Space}
\label{sec:interaction-space}
    \begin{figure*}[h]
        \includegraphics[width=\textwidth]{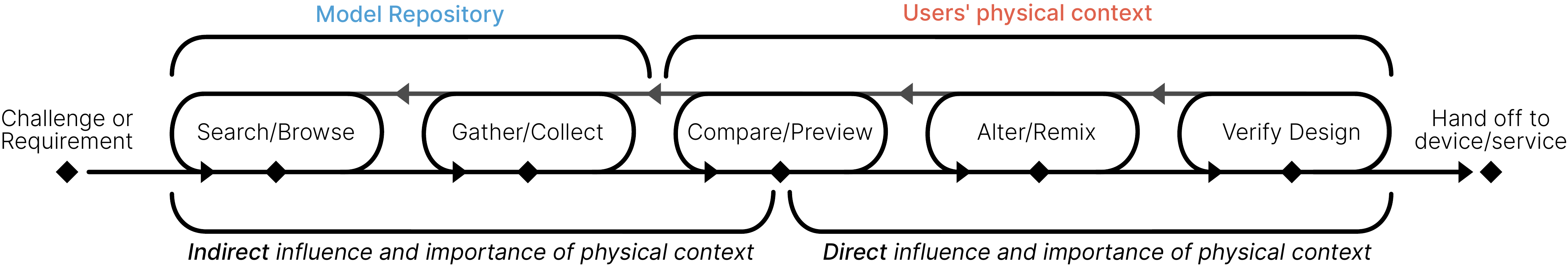}
        \caption{
            Steps which the concept behind \system emphasizes during the design process. 
            While each step's results feed in to the following ones, users always have the possibility to circle back to prior steps, for instance to refine their search terms, or gather more alternative solutions to use or remix.
        }
        \label{fig:process}
    \end{figure*}
    \system aims to allow users to omit the process of ''hand-crafting'' (e.g., 3D-modelling) while still retaining \emph{meaningful} alterations and customizations with respect to the users' physical context.
    This is what we specify as \emph{in-situ} \paradigm.
    There, 2 aspects of interaction are important: \emph{functional} interaction and \emph{aesthetic} interaction.
    \emph{Functional interaction} describes the actual practical tasks a fabricated artifact may fulfill.
    For example, whether a mount for a phone is actually able to hold it in a desired position, or whether a hook is mounted low enough to be reached and simultaneously high enough so that the clothing suspended from it does not touch the floor. 
    These constraints and requirements are often found only in the specific users physical context and emerge from the spatial configuration their context has.
    While some constraints may be reduced to standardized components, they can rarely provide a complete picture of the environment an artifact will reside and function in.
    \emph{Aesthetic interaction} describes the visual level of interaction between newly fabricated artifact and its future environment.
    This can be based on personal judgment of design, design consistency and general visual appeal.
    For instance, a newly acquired decorative planter may or may not fit the remaining objects on the countertop it is meant to be placed on.
    To ensure both appropriate \emph{aesthetic} and \emph{functional interaction}, different directions can be taken by users.
    They may rely either on measuring or on coarse visual judgement.
    They may also either accept the first adequate solution, or iterate further, either out of pure desire to do so, or if the first iteration does not fit its purpose\footnote{this excludes failures during the fabrication process, which are still a relevant factor~\cite{hudsonUnderstandingNewcomers3D2016}.}.

    The interaction space surrounding \systemEmph and the \emph{in-situ} \paradigm paradigm consists of three fundamental principles:
    \begin{enumerate}%
        \item \textbf{Outsourced design} effort, relying on existing designs
        \begin{enumerate}\itemsep0em 
            \item Existing designs are found in the \textbf{real} world.
            \item Existing designs are found in the \textbf{virtual} world.
        \end{enumerate}
        \item \textbf{In-situ adaptation} effort and remixing, referencing the physical context
        \item Variable \textbf{degrees of effort} to reach one's goals to accommodate for different users and requirements
    \end{enumerate}
    
    With \system, we aim to (mostly) omit modelling from the process of personal fabrication, while retaining the potential benefits of a modelled artifact: the prospect of an ideally tailored solution.
    This is in line with the notion of personal design, in contrast to personal fabrication~\cite{borchersPersonalDesignManifesto2013,borchersInternetCustomMadeThings2013}, which abstracts from the specifics of \emph{manufacturing} and focuses on user-centered \emph{design processes}.
    However, we argue that neither design nor fabrication have to be local (i.e., happen at the location of and be carried out by the user) to provide a successful and tailored artifact that fulfills the users' requirements.
    Merely the successful configuration and tailoring of a solution, likely to exist in the diverse model repositories that have emerged, is a relevant and \emph{inherently personal} part of personal fabrication.
    For instance, a user will likely find a design for a broom holder online, and would merely have to configure its diameter -- if deemed necessary -- for it to be an adequate solution.
    It is then not relevant who designed it or who will fabricate it; merely the tailoring to the user's requirements is crucial.
    
    Figure \ref{fig:process} describes the conceptual process we propose for the \emph{in-situ} \paradigm paradigm behind \system.
    While a similar notion already exists when one considers model repositories, we emphasize the unification of remote model repositories and the users' physical context as sources for artifacts at the location where they are meant to be employed.
    This is depicted in table \ref{tab:origins}, in combination with two distinct patterns of (re-)use: ''as intended'' in contrast to ''remixed / misued''.
    Artifacts can be copied from the digital repository, or from existing objects in the users' vicinity and either be used according to their original specification (with simple alterations like scaling), or be creatively misused (e.g., repurposing a decorative figure to serve as a phone mount).
    
    \begin{table}[b!]
        \includegraphics[width=\linewidth]{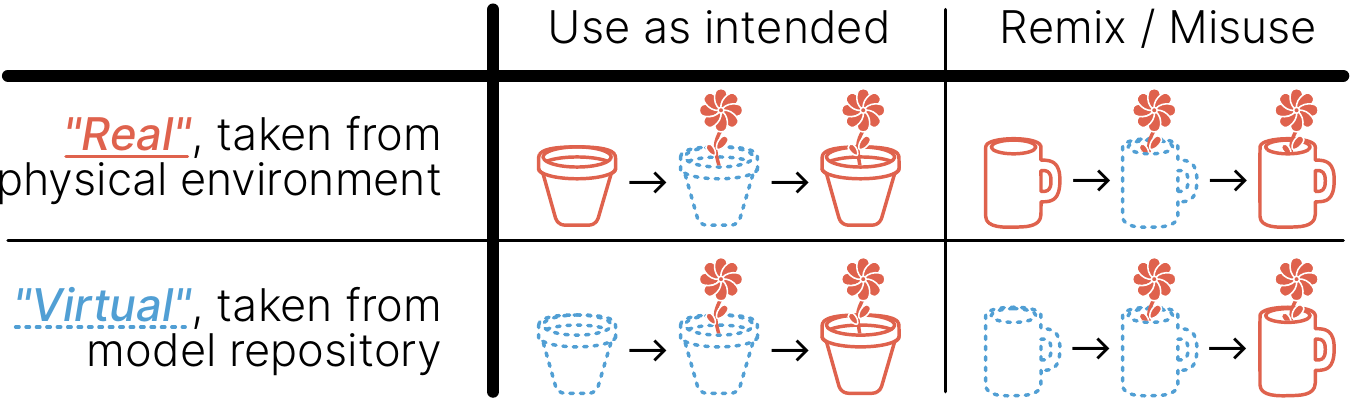}
        \caption{
            Model origins (model repository, physical environment) combined with two distinct usage patterns: largely unaltered use (i.e.,  ''as intended'') and remixed use or misuse. 
            Procedure per cell: retrieval, in-situ preview, fabricated result.
        }
        \label{tab:origins}
    \end{table}
    
    Starting with a specific goal, requirement or desire, the users initiate a search in the repository, or browse it without a clear search term.
    The users then may start gathering fitting alternatives for the task at hand.
    Up until this point, the interaction with \system is comparable to one with a model repository.
    The influence of the physical context is indirect, as it described the prior requirements.
    In addition, \system treats the physical environment as an equitable model repository.
    The users then may start to compare their alternatives (e.g., a set of headphone stands). 
    With \system, this happens in-situ -- right at the location where the artifact will interact with its environment.
    This allows both visual (i.e., aesthetic) and, to a degree, functional judgement.
    Afterwards, the users may start to alter the artifact or remix it with the help of features found in their physical environment (e.g., the thickness of a shelf, or the diameter of a pot).
    After verifying the design's functionality visually (e.g., by checking clearances or diameters), the users then may hand off the design to be fabricated.
    Whether this happens in their own homes (e.g., with their own 3D-printer) or is outsourced (e.g., to a printing service) is less relevant, as the fabrication of the artifact is not an inherently personal part of the process.

    \system emphasizes \emph{outsourced design effort} by leveraging model repositories, while allowing users to preview and adapt the artifacts retrieved.
    Ideally, this allows users to omit the modelling process entirely.
    Consistently omitting modelling is a na\"{i}ve ambition.
    It may be valid if the user chooses to fabricate a fully standardized component (e.g., an M2 screw with 2cm length).
    Few problems that are addressable with the means of personal fabrication are truly unique and may have been solved by someone else.
    However, the constraints and specifics imposed by the users and their physical context are not as easy to outsource.
    Therefore, while \emph{modelling} from scratch might not be always needed, \emph{configuring} may suffice.
    This is offered by customizer tools, for instance by \tv\footnote{\url{www.thingiverse.com/apps/customizer}, Accessed: 2.9.19} or \mmf\footnote{\url{www.myminifactory.com/customize}, Accessed: 14.9.19}, where dimensions of explicitly parametrized designs can be freely altered.
    Personal fabrication's outlook is that each end every user is able to create custom-made, tailored artifacts for their very personal use case and context.
    In contrast to store-bought solutions, solutions that emerge from personal design and fabrication may achieve a high degree of fit and tailoring with respect to the users and their requirements.
    This does not necessarily mean that the design or the fabrication process need to happen ''from scratch'' and be done by the user.
    Ideally, only \emph{relevant} effort has to be spent by the user (while still being free to invest more time into it).
    With \system, we want to extend the notion of a model repository to any user's physical context, outsourcing any effort not inherently vital to address a requirement.

\section{Prototype Implementation}
\label{sec:prototype}
Our prototype system is implemented using Unity 2019.2 and the Magic Leap ML1\footnote{\url{www.magicleap.com/magic-leap-one}, Accessed: 14.9.19} head-mounted display (HMD). 
\system aims to be provide as much functionality as possible within a single system -- ideally to replace software like CAD, a slicer and a printer interface. %
The following sections describe the implementation of the system.
    
    \subsection{Architecture}
    The architecture of the system is centered around the Magic Leap HMD, along with the REST (REpresentational State Transfer) API to a model repository.
    As a data source, we chose \mmf instead of \tv, primarily because the former provides vetted and moderated results.
    Furthermore, \mmf emphasizes quality and printability of the provided models.
    The downside of this is a less abundant choice of models.
    Moreover, our search functionality explicitly filters out any results that do not permit remixing.
    An alteration for personal use only would likely comply with most licenses used in model repositories.
    It is nevertheless reasonable to feed the remixed models back into the ecosystem, if the users deem it appropriate.
    This is likely the case for adaptations that generalize to a degree, like addition of standardized mounts/fixtures or remixes that resulted from combination of multiple artifacts from the repository~\cite{oehlbergPatternsPhysicalDesign2015}.

    \subsection{Interaction}
        The interaction with the system is meant to provide the most relevant functions of a model repository interface, while combining them with the scene understanding and spatial visualization a mixed reality headset provides.
        This is meant to support the \paradigm-paradigm, by largely omitting modelling while retaining adaptivity of outsourced artifacts.
        This primarily includes searching the repository, choosing fitting models and previewing them.
        Figure \ref{fig:process} described the process users may follow with the \emph{in-situ} \paradigm paradigm.
        The following paragraphs describe the implementation of each step for \system.
        While they are described as a sequence, the users always have the option to return to prior steps to reevaluate their choices and the process (as seen in Figure \ref{fig:process}).
        The following figures were captured either with the help of the ''capture service'' of the Magic Leap HMD, or via ''Magic Leap Device Bridge'' (MLDB).
        All exhibit an offset between the augmented content and the physical environment.
        To the user of the HMD, the imagery is properly aligned with the environment and exhibits proper occlusion by the user's hands.

        \subsubsection{Searching and Gathering}
        \begin{figure}[h!]
            \includegraphics[width=0.495\linewidth]{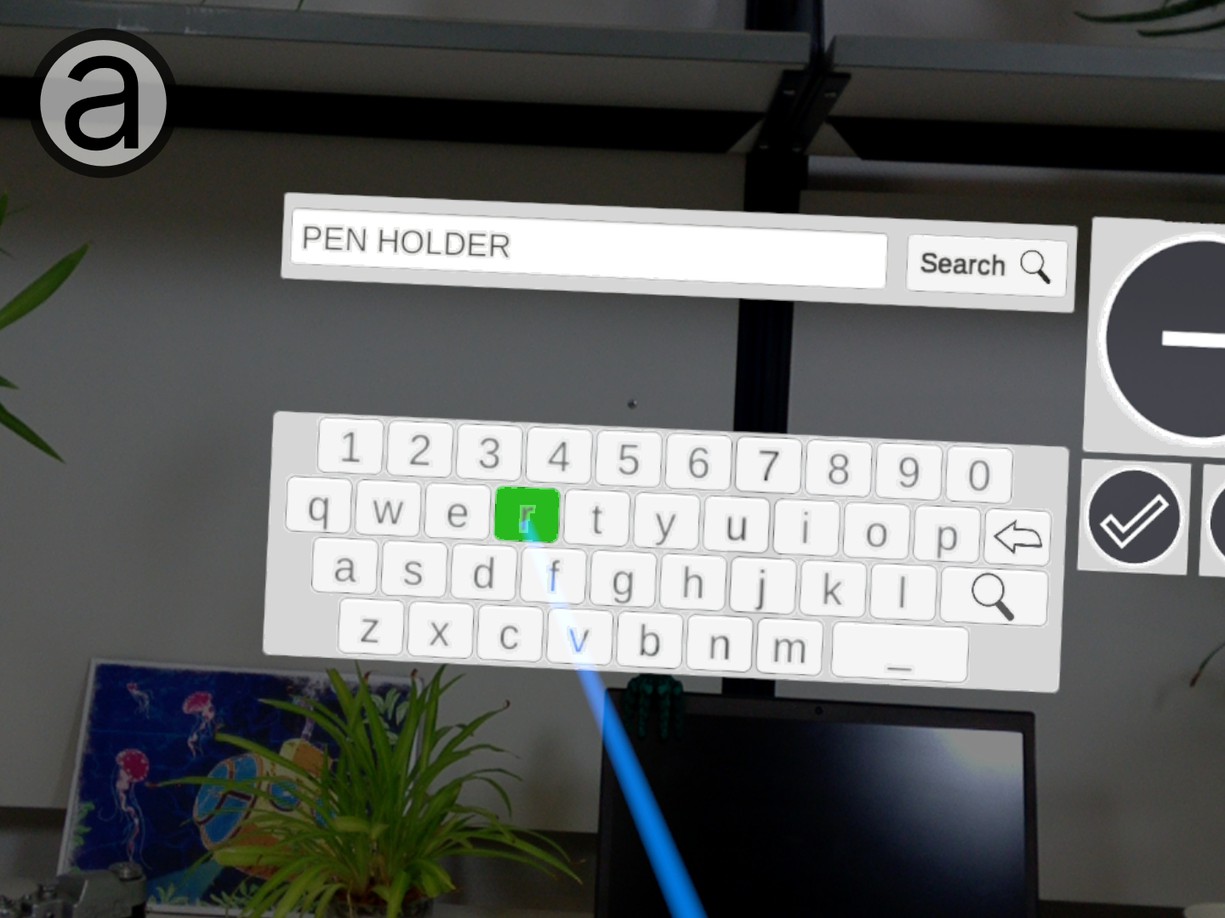}\hspace*{\fill}
            \includegraphics[width=0.495\linewidth]{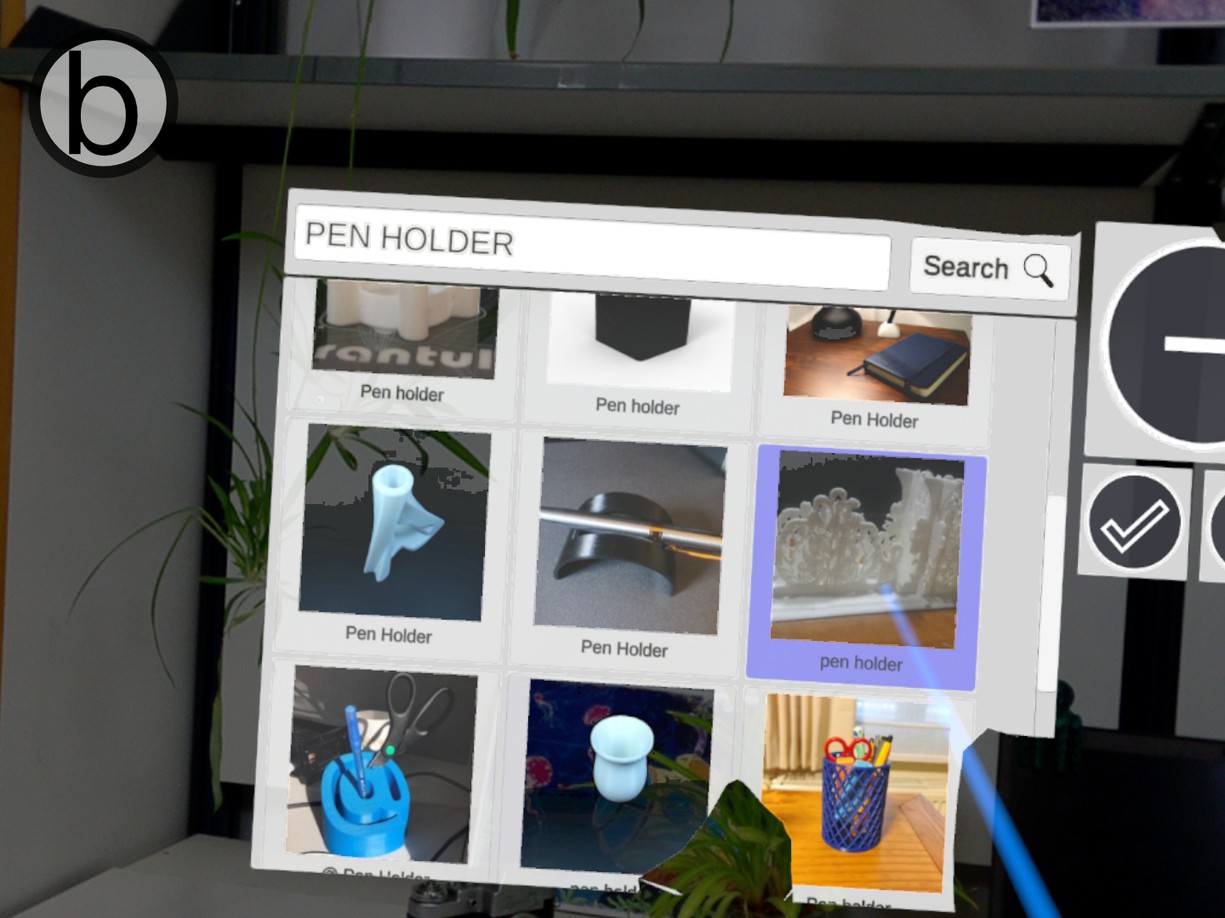}
            \caption{
                Initial interface to perform search requests.
                Users can enter their search terms (a), scroll through a set of previews of the results (b) and can minimize this UI when needed.
            }\label{fig:2dui}
        \end{figure}
        
        Each design process with \system begins with the search interface, presented to the users (Figure \ref{fig:2dui}).
        There, they are able to enter arbitrary search terms, similarly to the well-known web interfaces of \tv or \mmf.
        The application then relays the search via REST to the API of \mmf, which returns a JSON response, used to populate the list of results.
        After a successful search, the users may scroll through the results, with the title and a thumbnail image being present.
        Selection of a result enqueues it to be downloaded and added to the preview carousel described next.
        
        \subsubsection{Comparison and Previewing}
        \begin{figure}[h!]
            \includegraphics[width=0.495\linewidth]{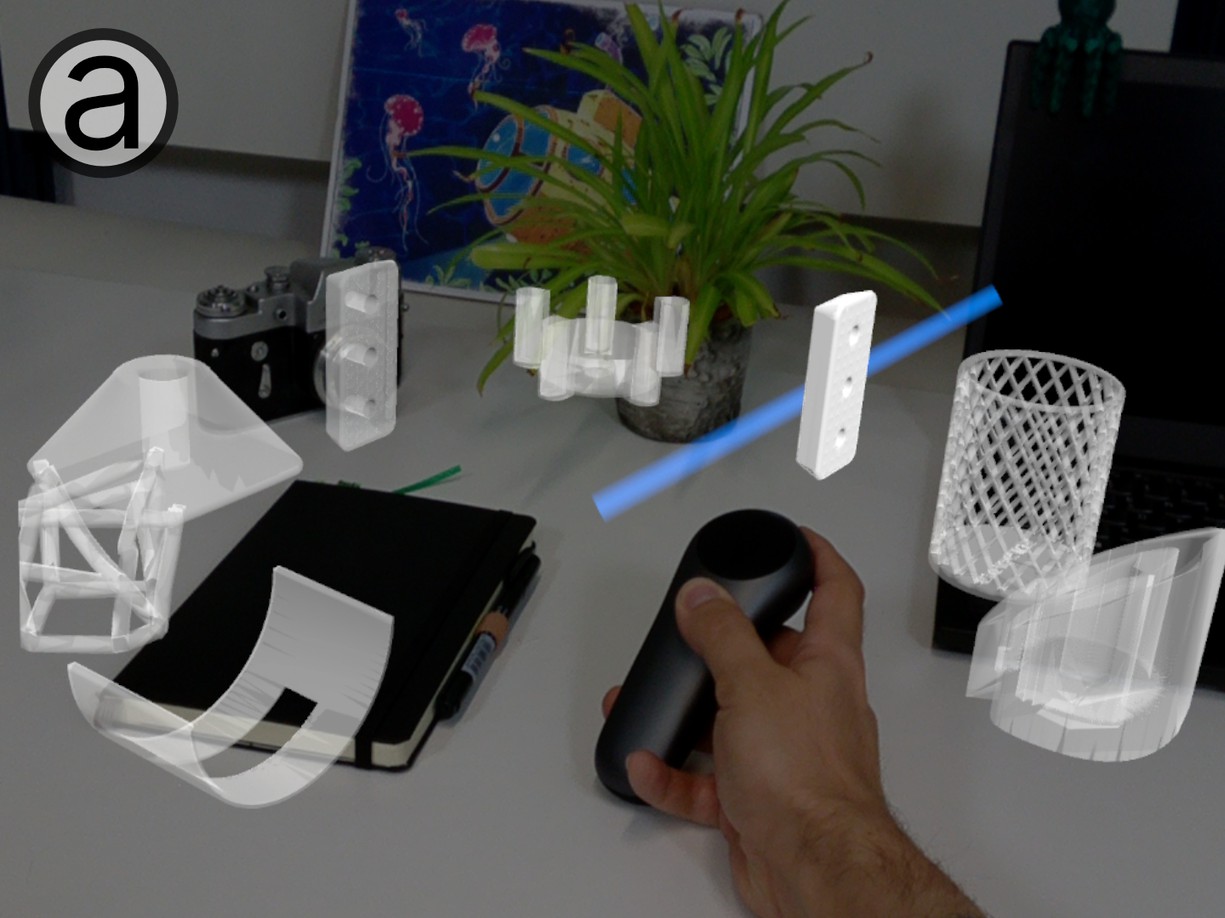}\hspace*{\fill}
            \includegraphics[width=0.495\linewidth]{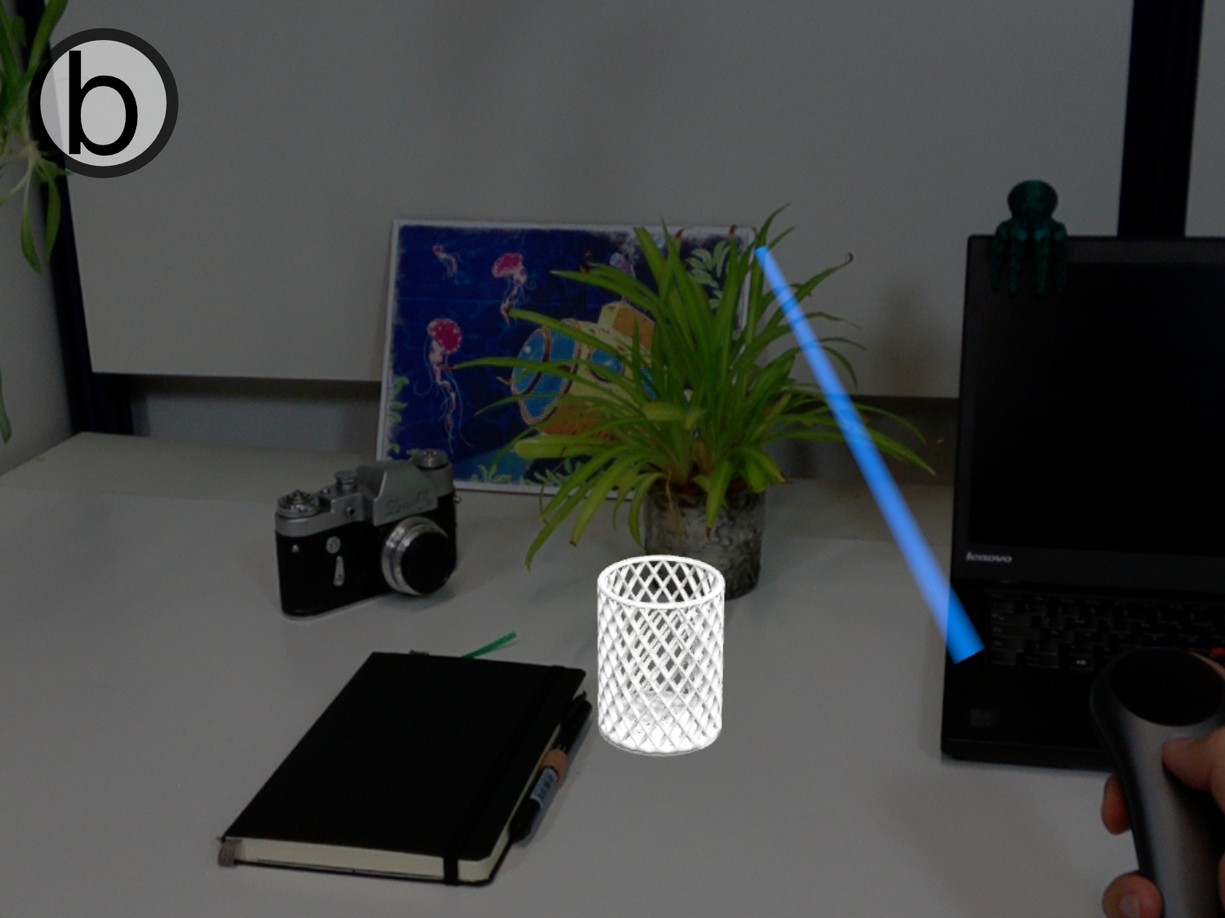}
            \caption{
                The placement carousel, which gathers all previously downloaded models (a).
                Users can cycle through the objects and place them in their environment (b).
            }\label{fig:carousel}
        \end{figure}
        
        Having collected an initial set of artifacts, the users may start to compare them in more detail.
        Each result is available through a carousel, arranged around and affixed to the controller, and is cycled through via the touchpad (Figure \ref{fig:carousel}, a).
        In contrast to the interface in the searching step, the users now gain insight into the spatial aspects of the model they have downloaded.
        They are now able to examine the entire geometry to judge the functionality or the appeal of the artifact.
        By holding the carousel where the artifact is meant to be employed and cycling through the options, users may directly compare their available alternatives.
        Artifacts that do not meet their requirements can be removed from the list of options.
        The models can be placed and affixed into the space around the user (Figure \ref{fig:carousel}, b).
        Each of the aforementioned actions is further supported by haptic and visual feedback.
        This allows the user to interact further with them, as described in the next section.
        
        \subsubsection{Alteration and Remixing}
            \begin{figure}[h!]
                \includegraphics[width=0.495\linewidth]{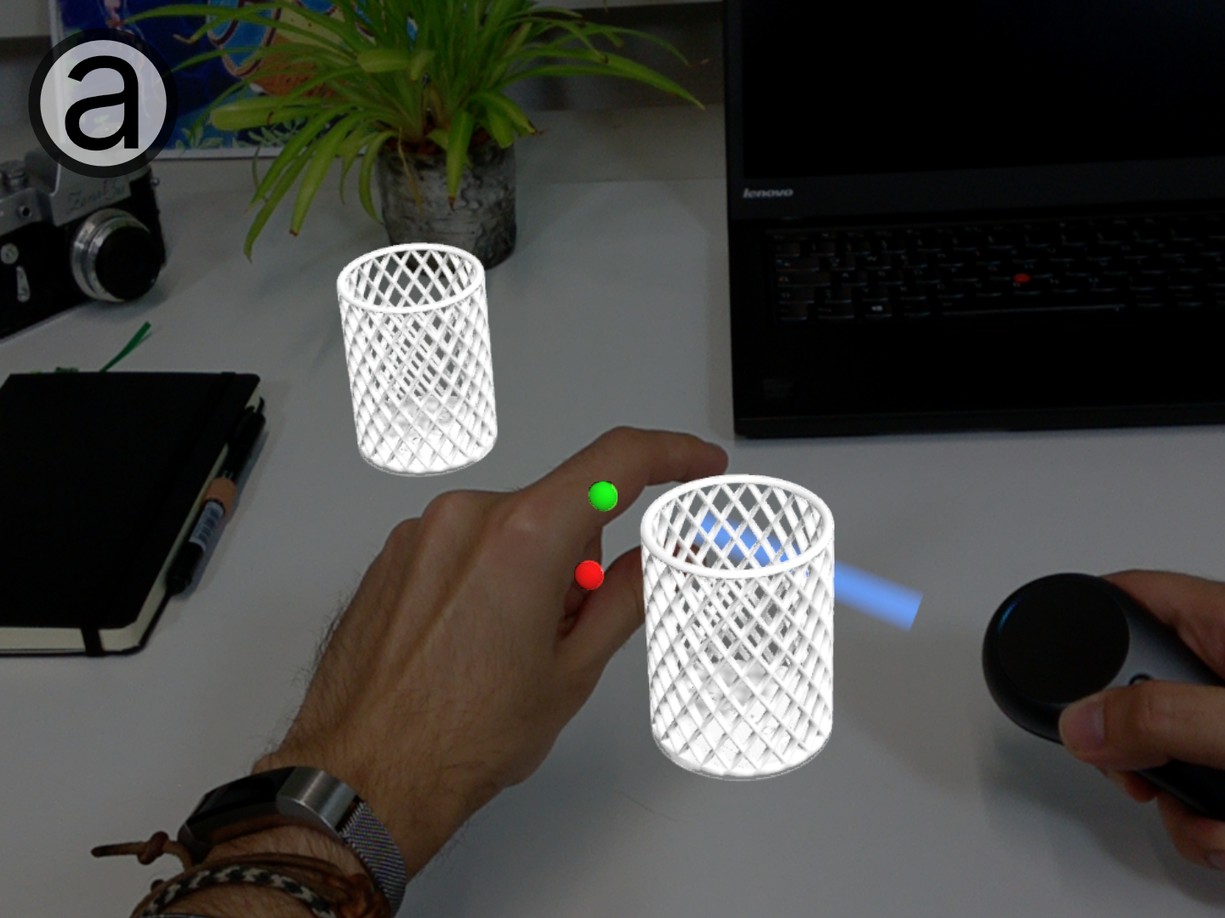}\hspace*{\fill}
                \includegraphics[width=0.495\linewidth]{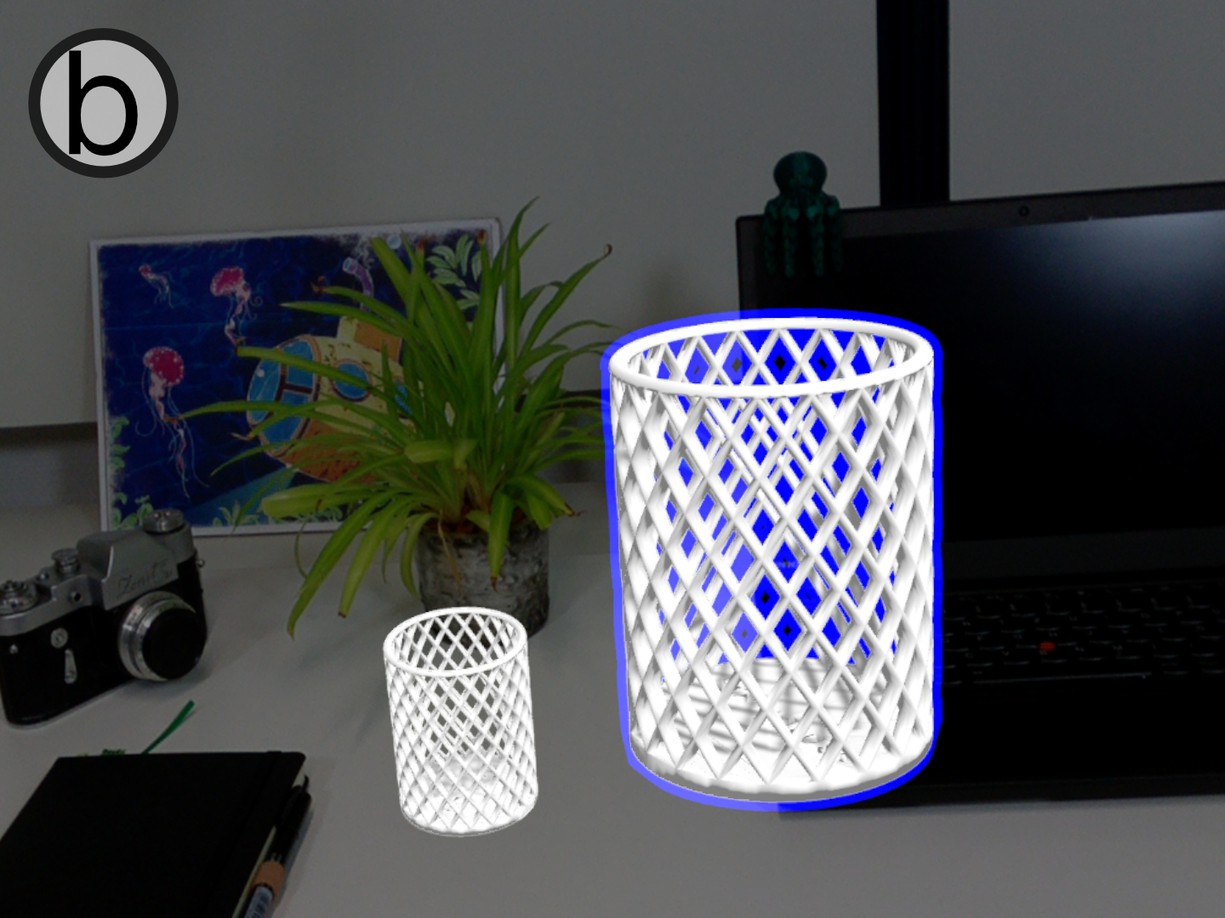}
                
                \caption{
                    Users can grab and move or rotate objects they have retrieved and placed. Selected objects are highlighted with an outline (b).
                    To scale them, users grab the object with the controller, and perform a pinch gesture (a).
                }\label{fig:grabmoveselect}
            \end{figure}
            Having placed an amount of models of their choosing, users can now interact and alter them in greater detail.
            The possible alterations include moving the object, rotating it and scaling (Figure \ref{fig:grabmoveselect}).
            To move or rotate an object, users grab it with their controller and directly manipulate it while holding the trigger button.
            Scaling also requires a ''grabbing'' with the controller -- additionally, users have to perform a pinch gesture with their other hand, while moving their hands apart.
            
            \begin{figure}[h!]
                \includegraphics[width=0.495\linewidth]{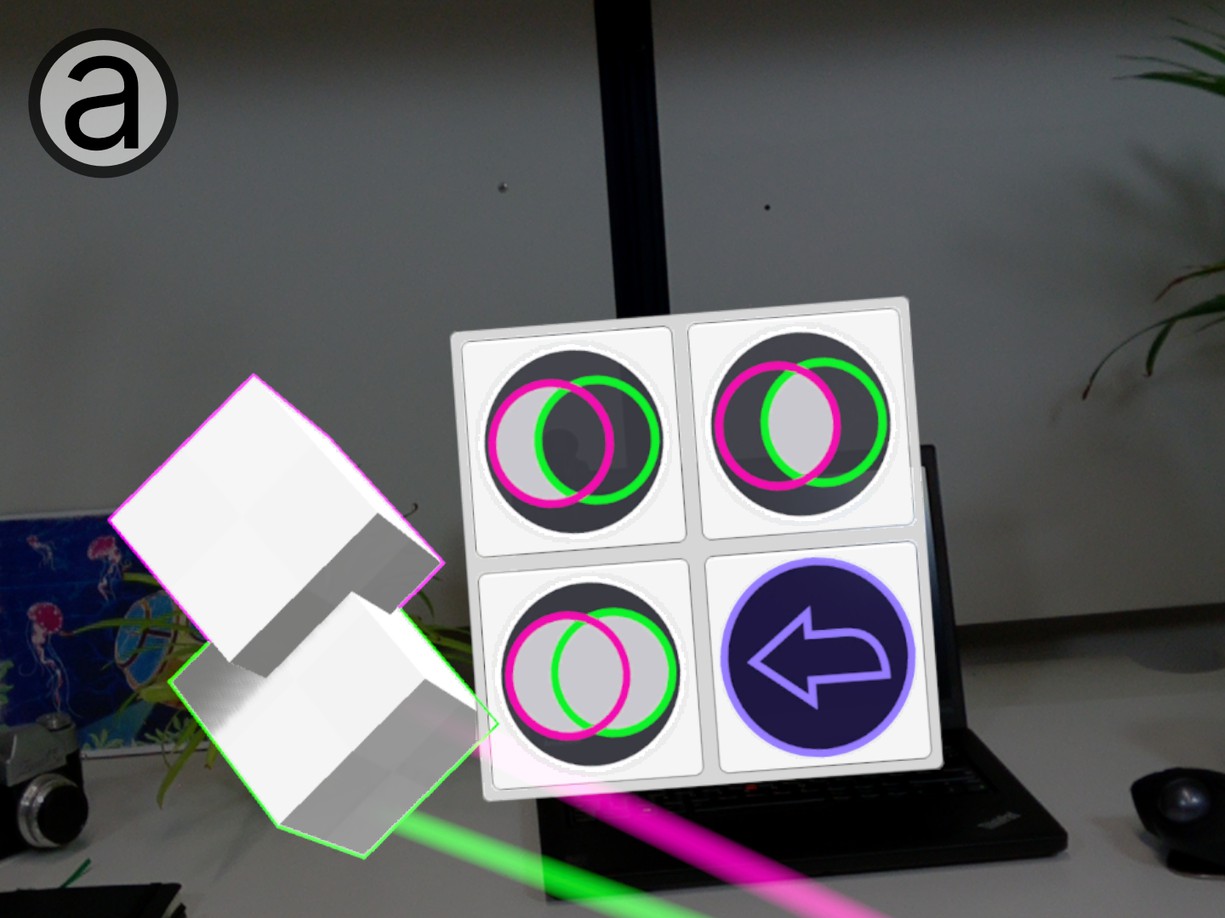}\hspace*{\fill}
                \includegraphics[width=0.495\linewidth]{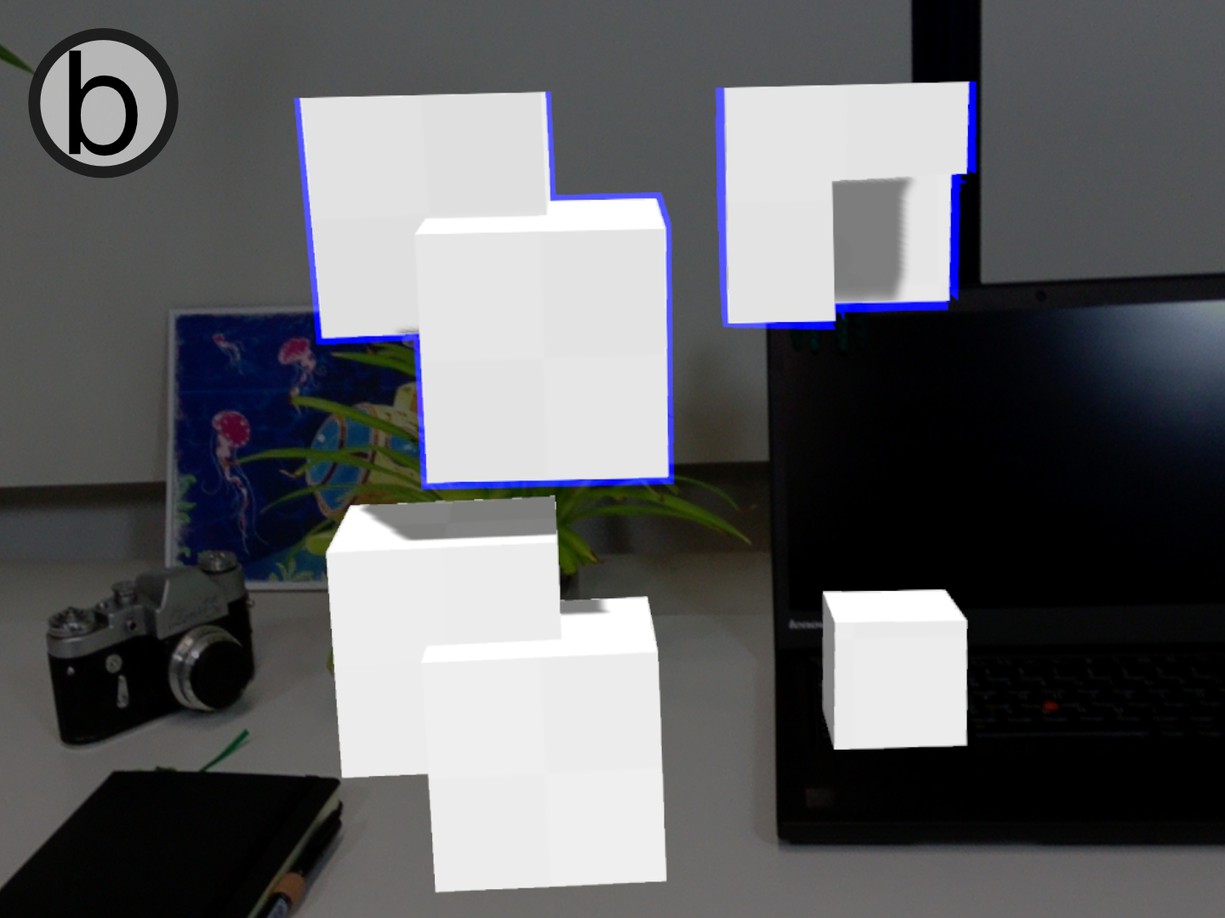}
                \caption{
                    Interface for CSG operations and their reversal (undo). As the subtract operation is not commutative, the selection is color-coded (a).
                    Results of all 3 operation types (b).
                }\label{fig:csg1}
            \end{figure}
            
\begin{figure}[h!]
                \includegraphics[width=0.495\linewidth]{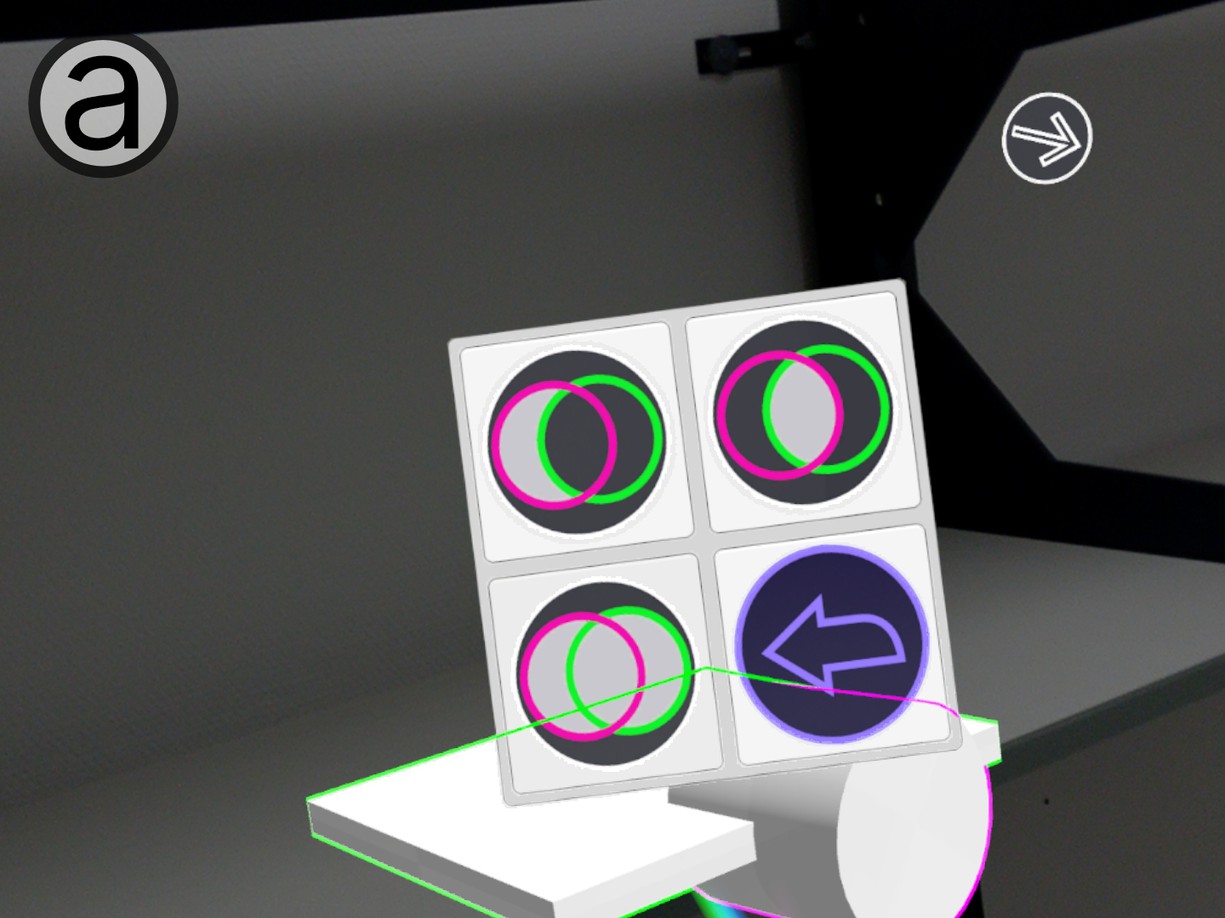}\hspace*{\fill}
                \includegraphics[width=0.495\linewidth]{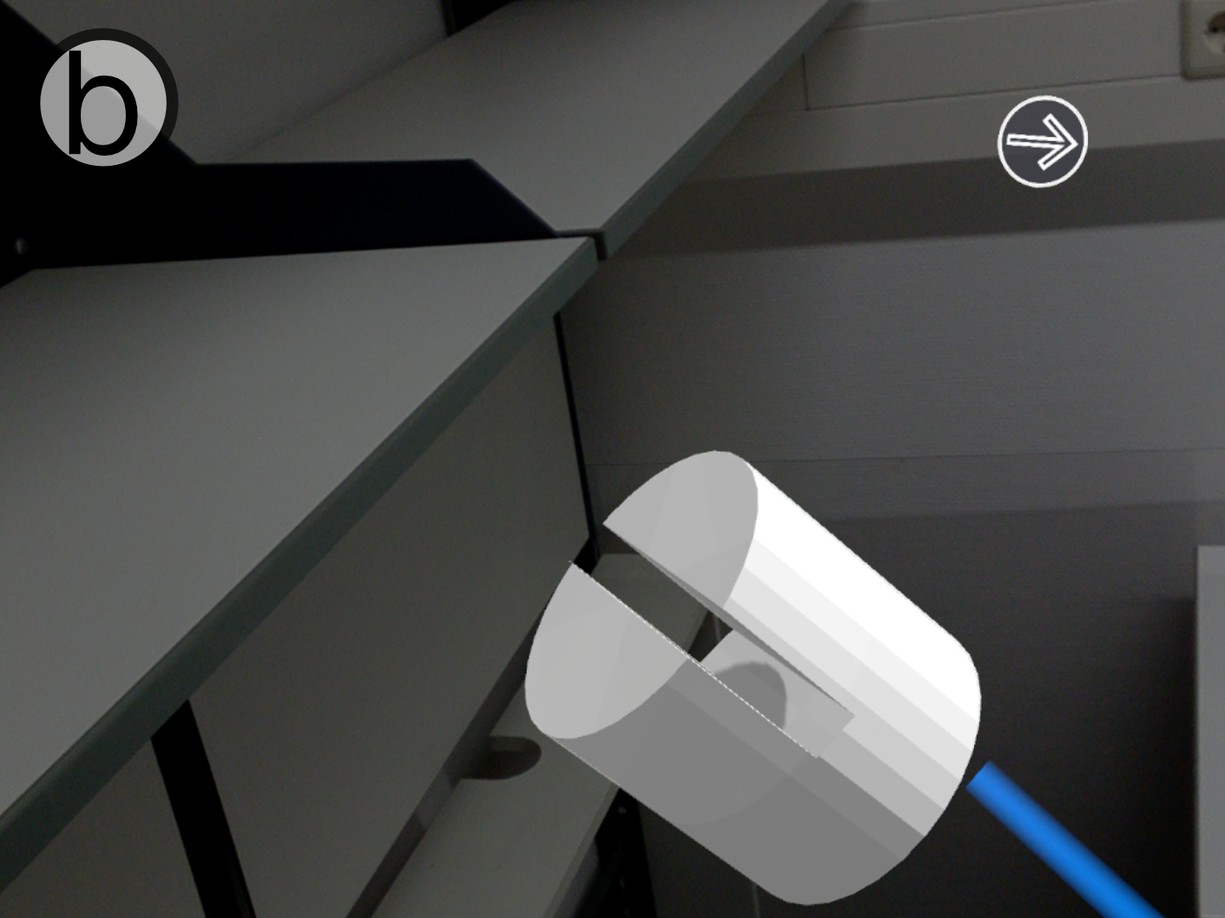}
                \caption{
                    CSG operations can likewise be based on real-world geometry, if available.
                    Selecting the shelf and the cylinder while they intersect (a), allows the user to subtract the shelf from the cylinder for a friction-fit (b).
                }\label{fig:csg2}
            \end{figure}

            Beyond these basic operations, \system provides Boolean operations (CSG, constructive solid geometry~\cite{requichaConstructiveSolidGeometry1977,laidlawConstructiveSolidGeometry1986}) for the placed models. 
            This allows users to combine (\emph{union}) models, subtract them from one another (\emph{difference}) and \emph{intersect} them. 
            After selecting two models, users are presented with an interface to choose one of the aforementioned operations (Figure \ref{fig:csg1}).
            Union also serves as a simple grouping feature, known from other applications.
            These operations are considered to be destructive, and can therefore be undone (Figure \ref{fig:csg1}, a)
            To allow users to alter the models they download, \system also provides access to 4 default primitives (cube, sphere, pyramid, cylinder), that can be interacted with, similarly to other models.
            They also serve as easy-to-use features for CSG Operations, where no suitable counterpart can be found in the users' physical environment or the model repository (Figure \ref{fig:csg2}).
            
            Our initial approach relied on the reconstructed environment mesh provided by the ML1 HMD.
            However, the resolution of the available mesh was too coarse to allow for \emph{precise} geometric interaction between artifacts and the environment.
            Likewise, treating the device like a 3D-scanner does not yield appropriate results yet.
            As most modern HMDs provide some degree of depth sensing and world reconstruction, we argue that with sufficient maturity of the devices, a detailed environment mesh can be made available to users.
            It then serves as an additional geometry to reference in the process of customization.
            In its current state, \system relies on marker tracking and thereby reproduces environment features in an appropriate fidelity.
            All ''copy and paste'' or CSG operations based on real-world artifacts or geometry are subsequently based on previously scanned or otherwise acquired 3D geometry.

    \subsection{Preprocessing, Postprocessing and Output}
\begin{figure}[h!]
            \includegraphics[width=\linewidth]{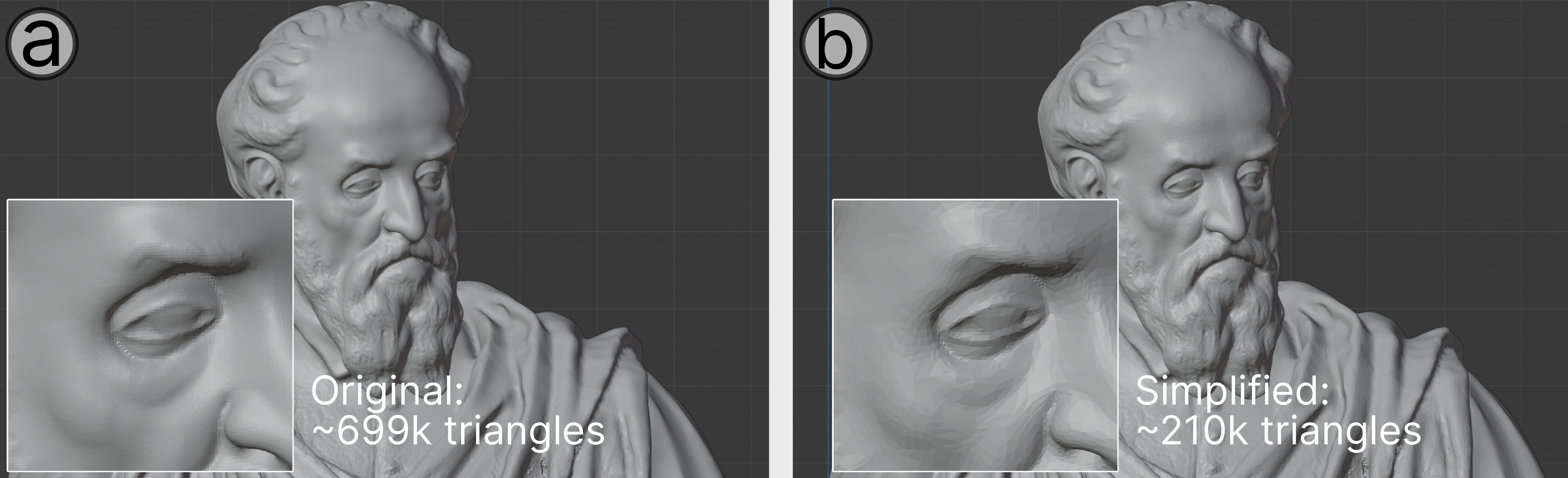}
            \caption{
               A detailed model before (a) and after (b) the applied simplification as exported by \system. The loss in quality is almost negligible.
            }\label{fig:simplified}
        \end{figure}

        Multiple stages of processing happen without user intervention.
        After the download of a model, the mesh is pre-processed, prior to being handed to the user to be altered.
        Depending on the amount of detail a mesh has, a simplification/decimation step is executed.
        This is particularly relevant for highly detailed models, like 3D-scanned sculptures.
        An example can be seen in Figure \ref{fig:simplified}, where a quality factor of 30\% is applied to the model, reducing the polygon count from approximately 699k triangles down to approximately 210k.
        For low-detail meshes, this step is skipped, to preserve all features of the design.
        Afterwards, inconsistencies in terms of bounds and normal alignment are corrected.
        
\begin{figure}[h!]
            \includegraphics[width=\linewidth]{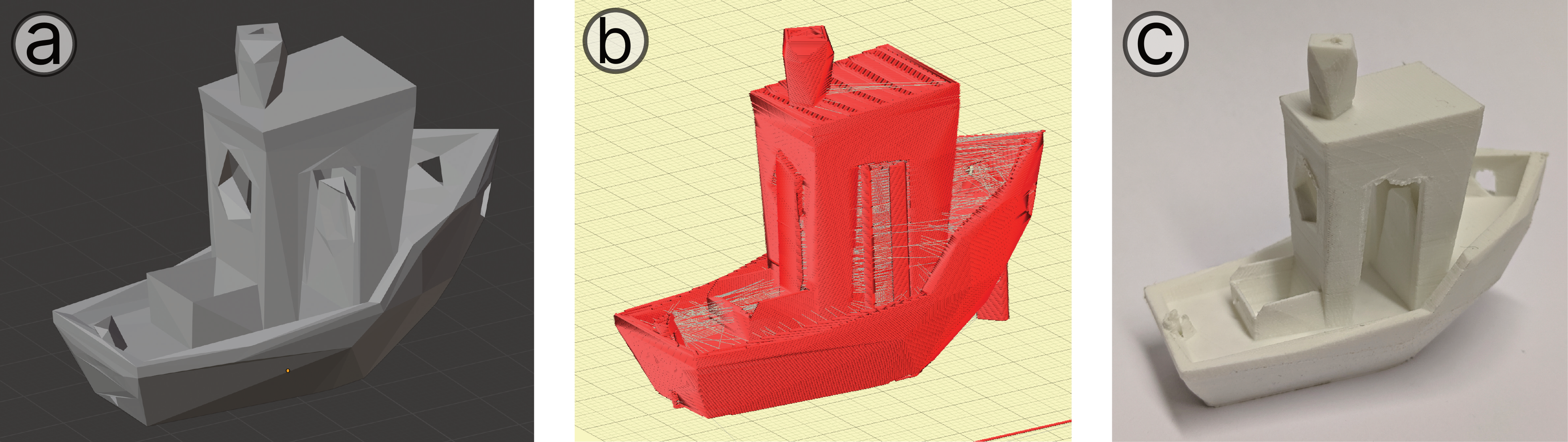}
            \caption[]{
                Exported mesh of a user-selected model (a).
                G-code generated on-device, based on this mesh, as visualized by Pronterface\protect\footnotemark~ (b). Printed result, with supports removed (c). 
            }\label{fig:stlgcode}
        \end{figure}

        After completing all necessary operations, the user may start to finalize the design.
        This is triggered by the save button on the interface, which initiates the output process.
        The design is then saved in .stl format to the local storage of the device. 
        As an additional step, \system can generate .gcode files directly on-device. 
        Using the gsSlicer\footnote{\url{www.github.com/gradientspace/gsSlicer}, Accessed: 12.9.19} library, a machine-readable description for the fabrication process is generated.
        The results (exported mesh and the .gcode generated from it), including the necessary support structures for 3D-printing, can be seen in Figure \ref{fig:stlgcode}.
    
        \footnotetext{\url{www.pronterface.com/}, Accessed: 10.9.19}

\section{Usage and Application Scenarios}
\label{sec:usage}
    The following paragraphs provide brief walkthroughs for 2 tasks users may tackle with \system.
    They aim to highlight the fact that despite providing only rudimentary modelling capabilities, \system allows for multiple, equally viable paths to a solution fulfilling the users' \emph{functional} and \emph{aesthetic requirements} alike.
    Each path either emphasizes the users' physical environment or the outsourced designs to a greater degree.
    Furthermore, each path exhibits a varying degree of effort that is needed to achieve a satisfying solution for the task.
    With these example scenarios, we want to emphasize the appeal of an in-situ \paradigm procedure in the users' own physical context.

    \subsection{Walkthrough 1: Finding a Fitting  Pot for a Houseplant}
        A user has recently acquired a small houseplant.
        He intends to replace the original planter with a more intricate one.
        The target artifact has to fulfill both aesthetic requirements (i.e., fit the theme of his desktop), and functional requirements (i.e., fit the inner pot's diameter).
        \system aims to support the user fulfill both requirements, offering variable degrees of effort needed.
        
        \subsubsection{Path 1 - Adapting a Fitting Design}
            \begin{figure}[h!]
                \includegraphics[width=0.328\linewidth]{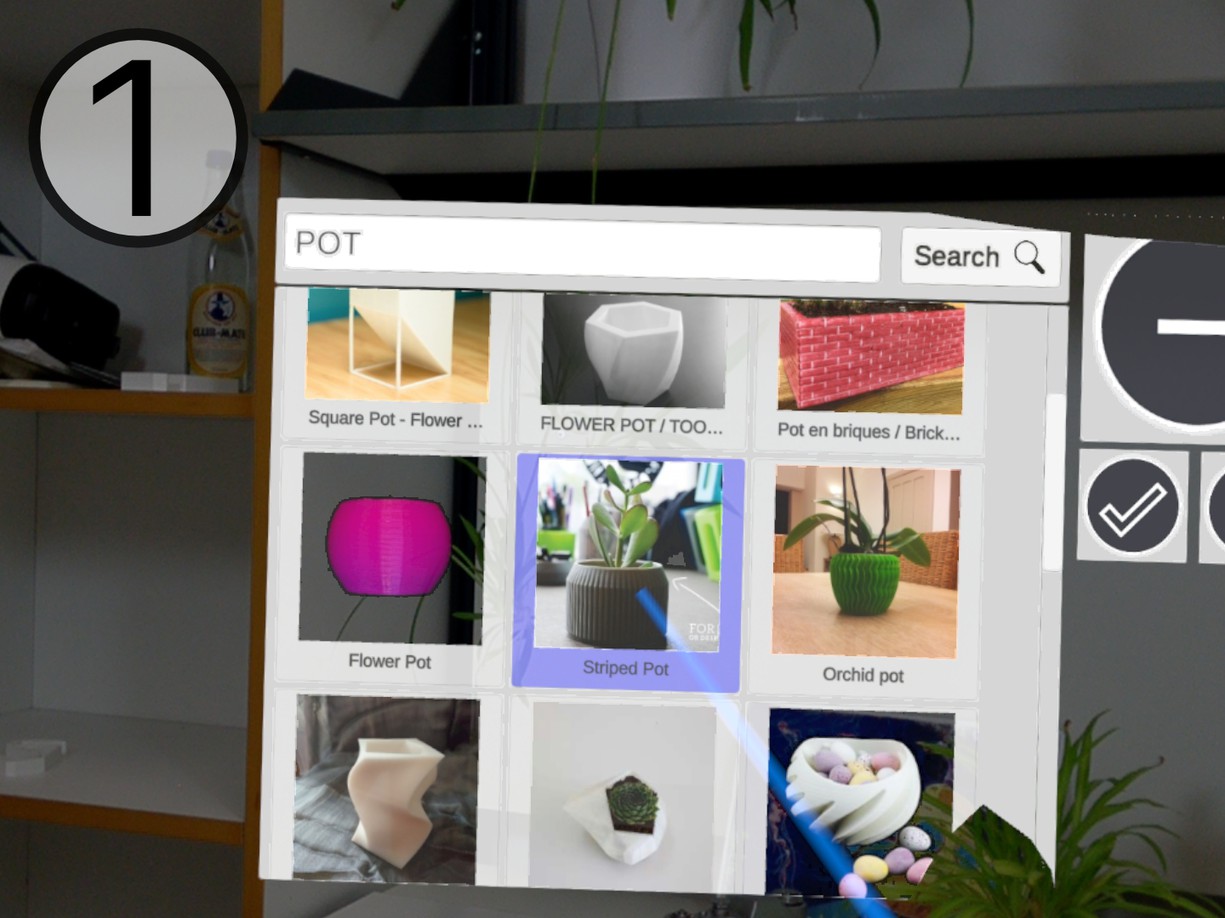}\hspace*{\fill}
                \includegraphics[width=0.328\linewidth]{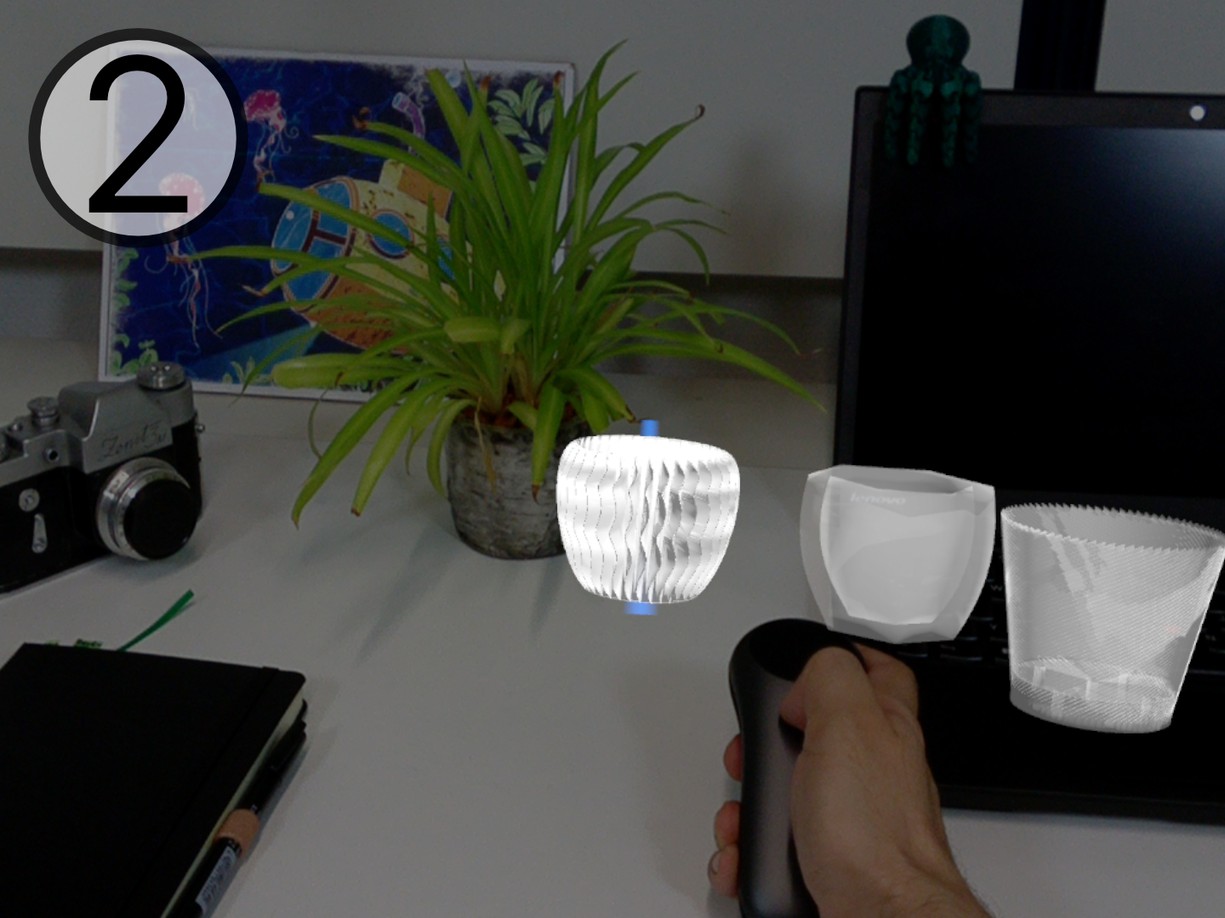}\hspace*{\fill}
                \includegraphics[width=0.328\linewidth]{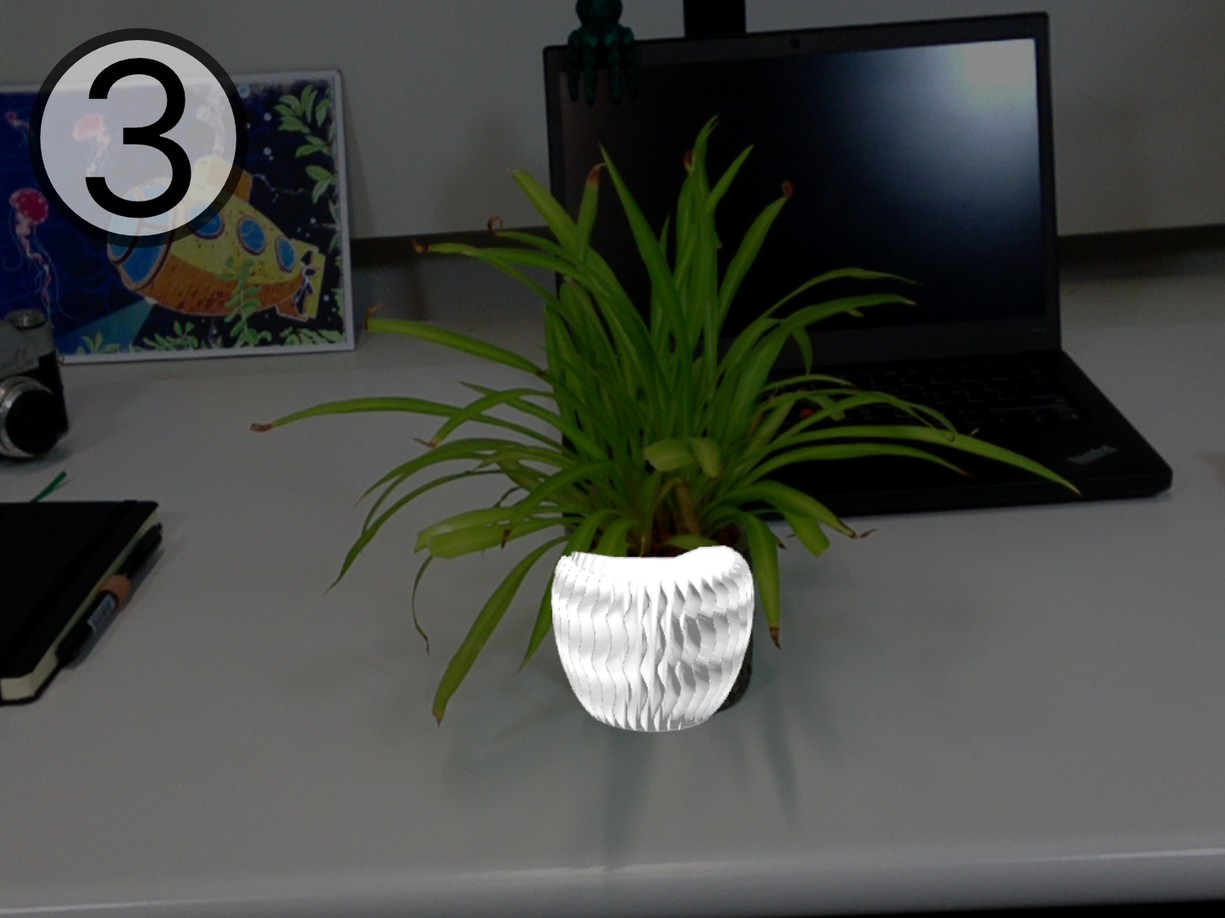}\hspace*{\fill}
                \caption{
                Adaptation of a fitting planter design to the existing plant.
                Browsing (1), comparing alternatives (2), customizing / scaling to fit (3).
                }
                \label{fig:pot-p-1}
            \end{figure}
            First, the user chooses to search for ''pot'', and initially selects a set of alternatives based on the thumbnails.
            He then cycles through the downloaded planters, removing the ones that do not appeal to him.
            Having decided on one design, he starts to scale the \emph{virtual} pot until it fits the diameter and the depth of the \emph{real} plant (Figure \ref{fig:pot-p-1}).
            If the scaled variant of the planter loses its visual appeal, the user may circle back to an earlier step, either searching and gathering more alternatives, or choosing a different one from the initially downloaded set.
            
        \subsubsection{Path 2 - Repurposing/Misusing/Remixing a Design}
            \begin{figure}[h!]
                \includegraphics[width=0.328\linewidth]{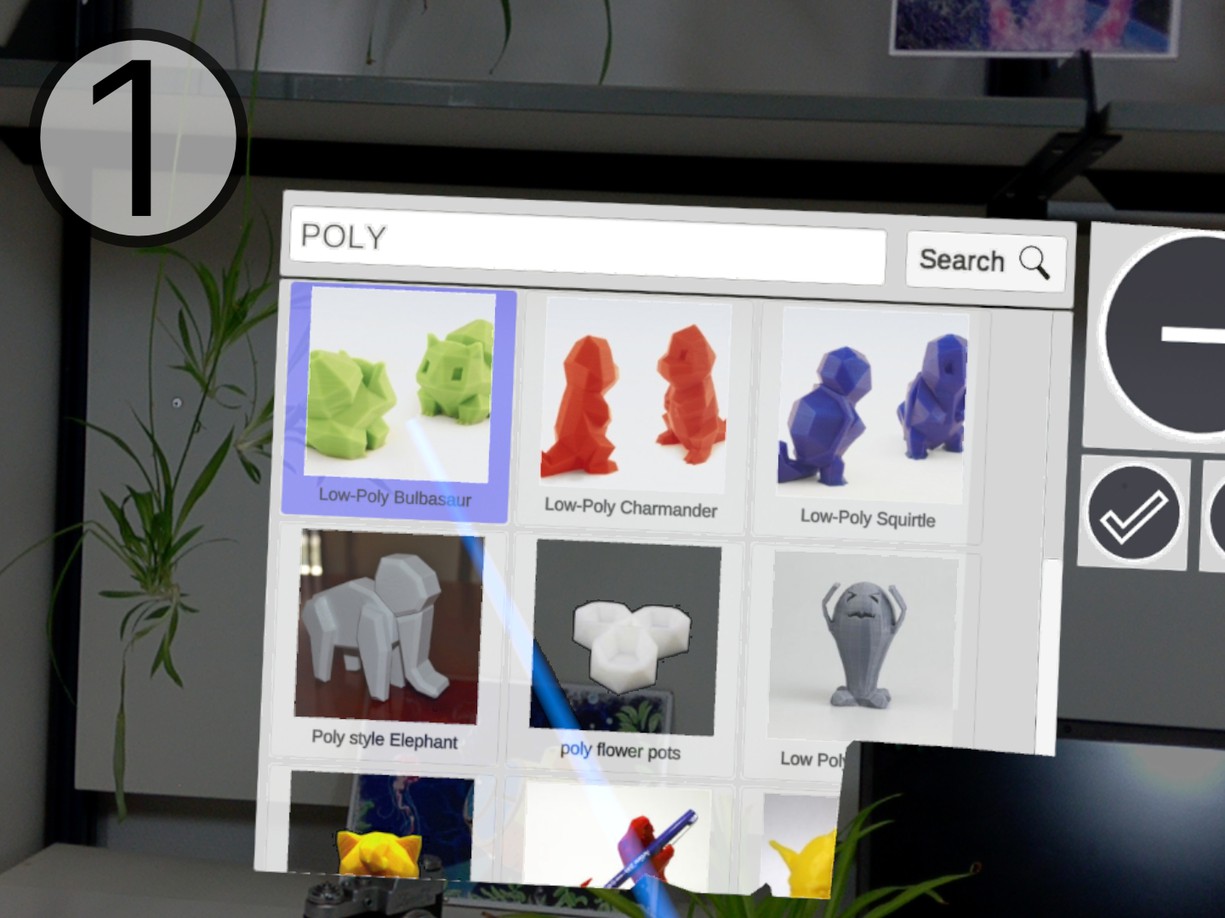}\hspace*{\fill}
                \includegraphics[width=0.328\linewidth]{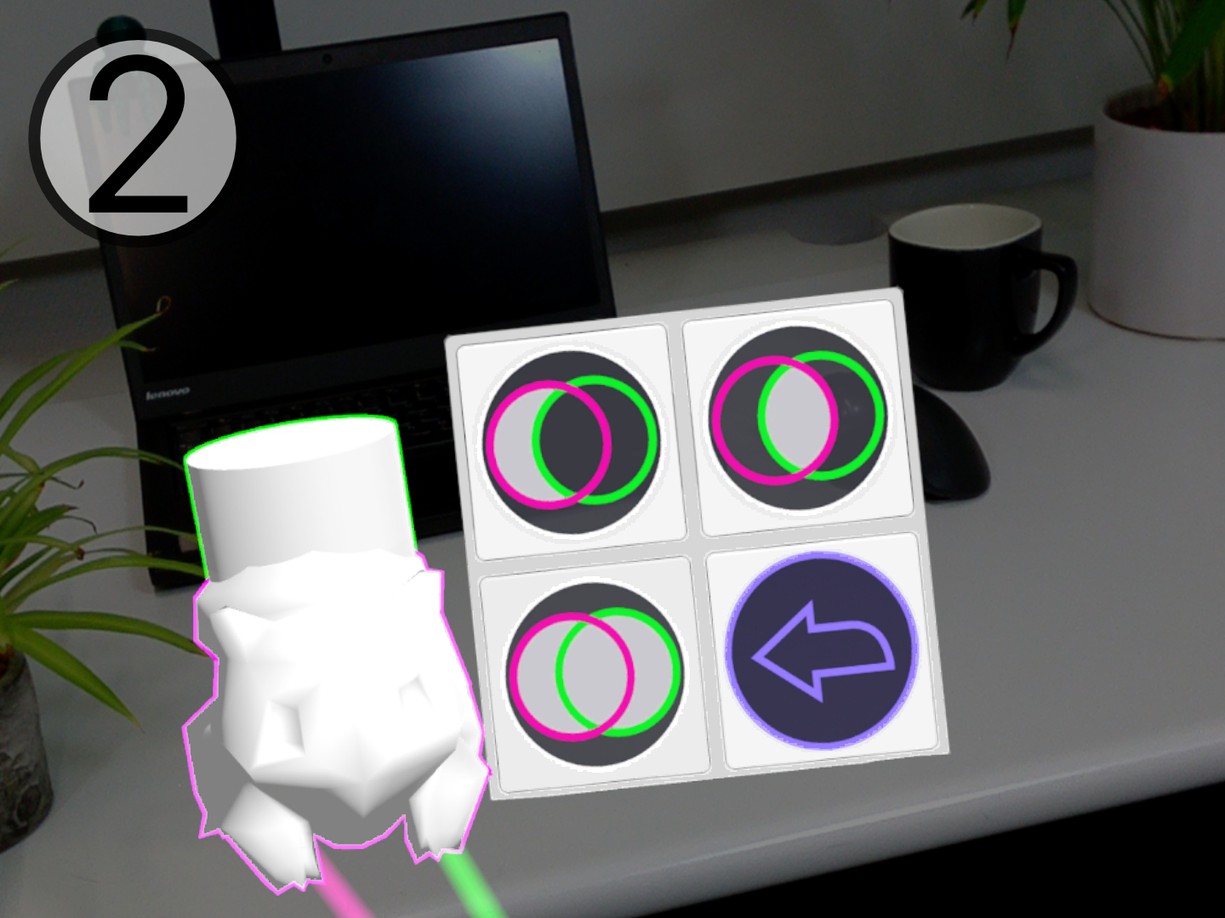}\hspace*{\fill}
                \includegraphics[width=0.328\linewidth]{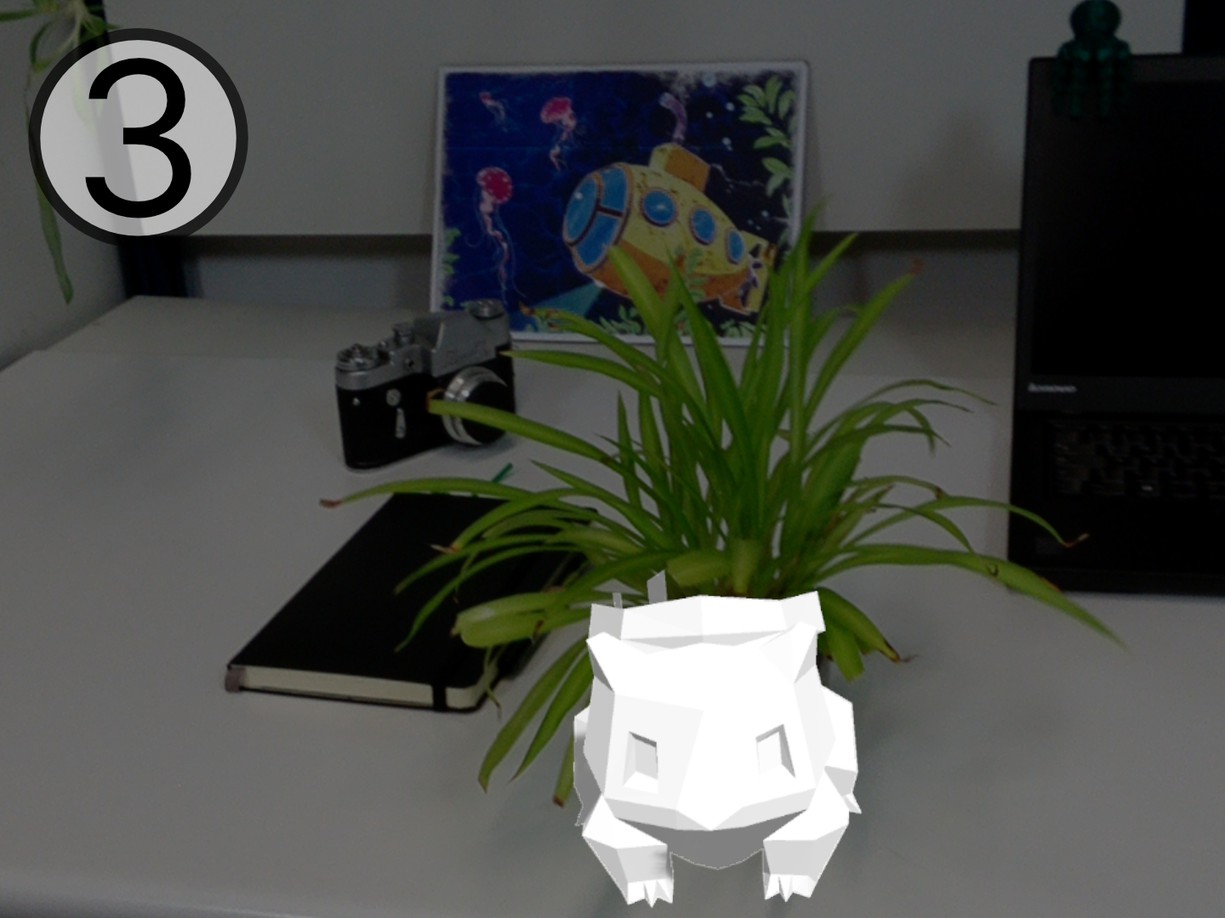}\hspace*{\fill}
                \caption{
                In-Situ remixing of a figure to create a planter.
                Searching for a base design (1), placing the components (figure and cylinder) and applying subtraction (2), visually verifying proportions (3).
                }
                \label{fig:pot-p-2}
            \end{figure}
            Instead of searching for a planter, the user instead aims to recreate a design where a plant's leaves represent a figures ''hairstyle'' (Figure \ref{fig:pot-p-2}).
            He downloads the figure, places it on his table and scales it to coarsely enclose the planter.
            Afterwards, he creates a cylinder primitive from the provided interface and moves it to intersect the figure.
            Applying the subtract CSG operation yields a hole for the planter to fit in.
            Alternatively, the subtractive part of the CSG operation may be the pot itself, if it is scanned in sufficient detail.
        
        \subsubsection{Path 3 - Replicating an Existing Artifact}
            \begin{figure}[h!]
                \includegraphics[width=0.328\linewidth]{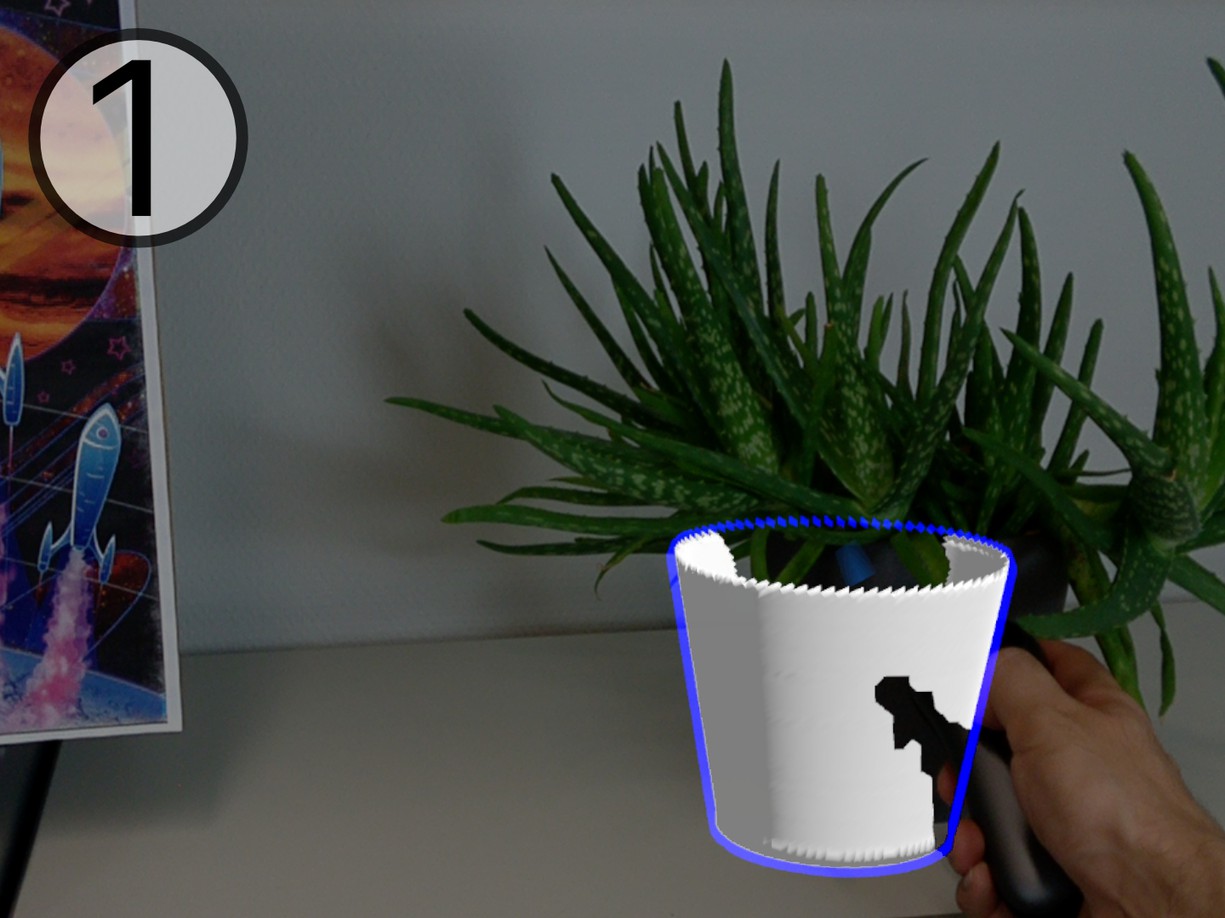}\hspace*{\fill}
                \includegraphics[width=0.328\linewidth]{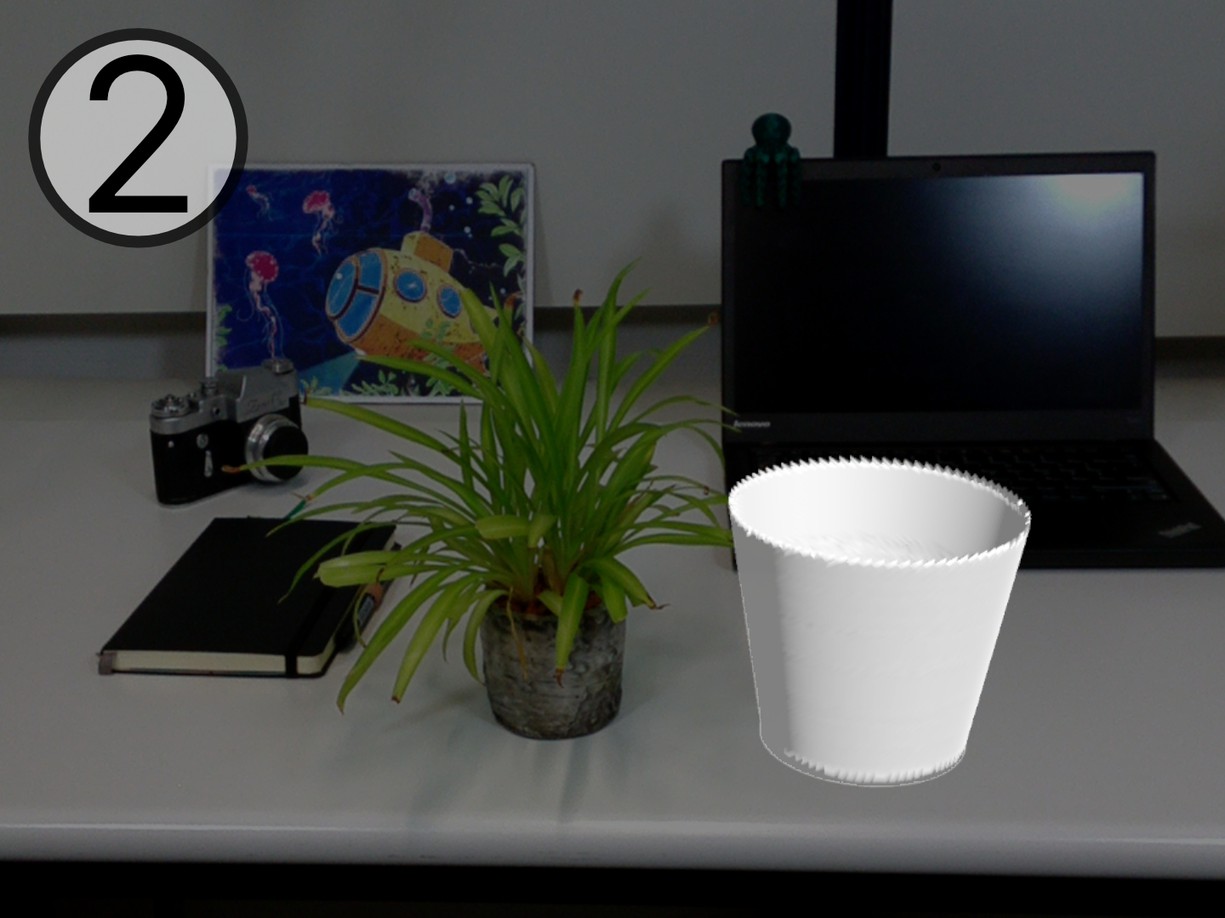}\hspace*{\fill}
                \includegraphics[width=0.328\linewidth]{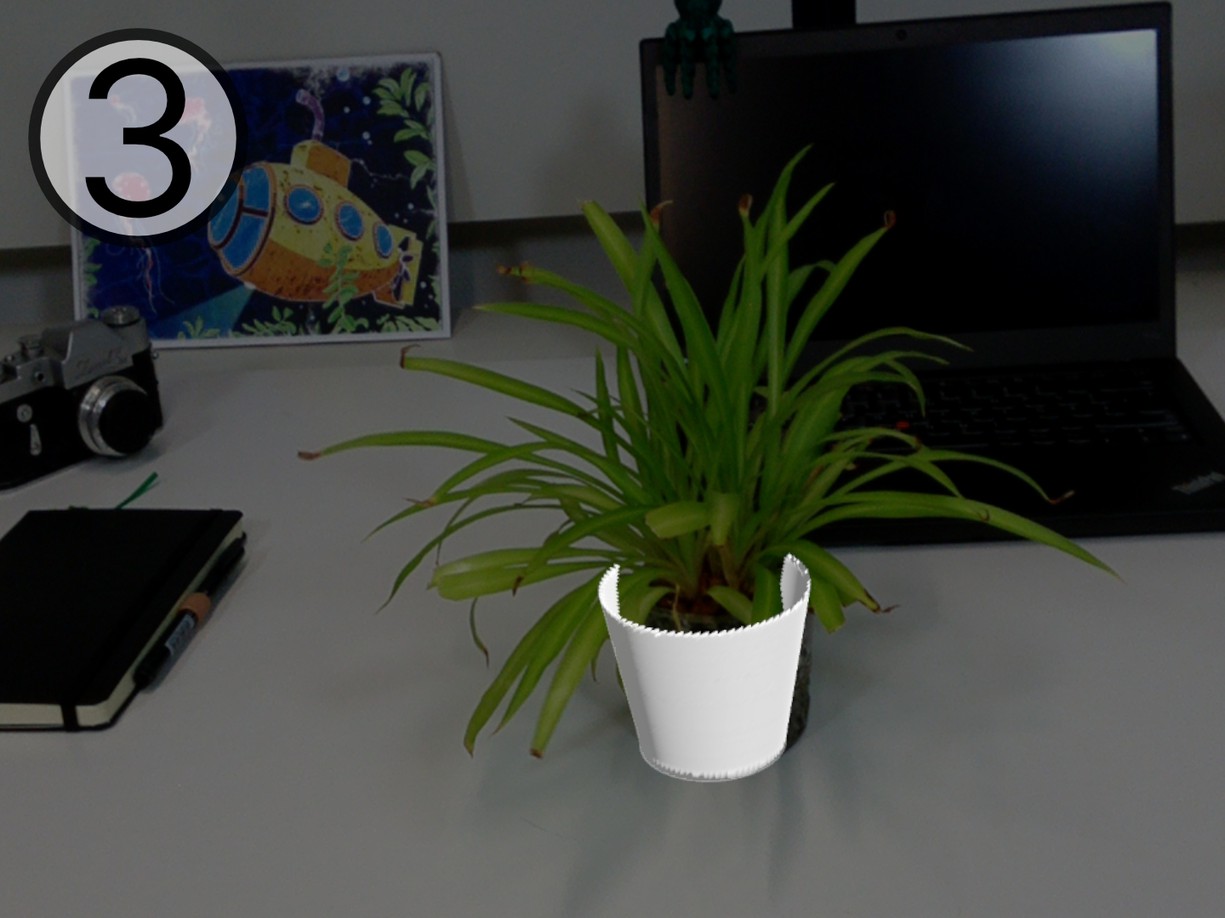}\hspace*{\fill}
                \caption{
                Replicating an existing physical artifact (planter).
                From left to right: Copying the mesh (1), previewing the result in terms of size (2), correcting the scale and proportions (3).
                }
                \label{fig:pot-p-3}
            \end{figure}
            It is also possible to employ a real-world ''copy-and-paste''-like procedure.
            The user may already have a planter in use that is both visually appealing and fulfills its function (Figure \ref{fig:pot-p-3}).
            Subsequently, there is no immediate need to start browsing for other designs.
            It suffices to select the existing planter, duplicate it and proceed with further alterations, if the need arises.
            If the mesh of the planter is not fully separated from the environment mesh, the user may place a cube primitive at the location, covering the object to be selected.
            The intersect CSG operation then would provide the user with a separate mesh.

\subsection{Walkthrough 2: Creating a Shelf-mounted Cloth Hook}
        For the second walkthrough, we present the task of finding and creating a cloth or coat hook, which is meant to be affixed on a shelf.
        The user is initially not sure, whether she wants to emphasize looks or functionality, and starts browsing the repository without a clearly defined path.
        As the repository presents a large amount of diverse artifacts, the user may feel compelled to repurpose or remix objects.
        
        \subsubsection{Path 1 - Adapting a Fitting Design}
            \begin{figure}[h!]
                \includegraphics[width=0.328\linewidth]{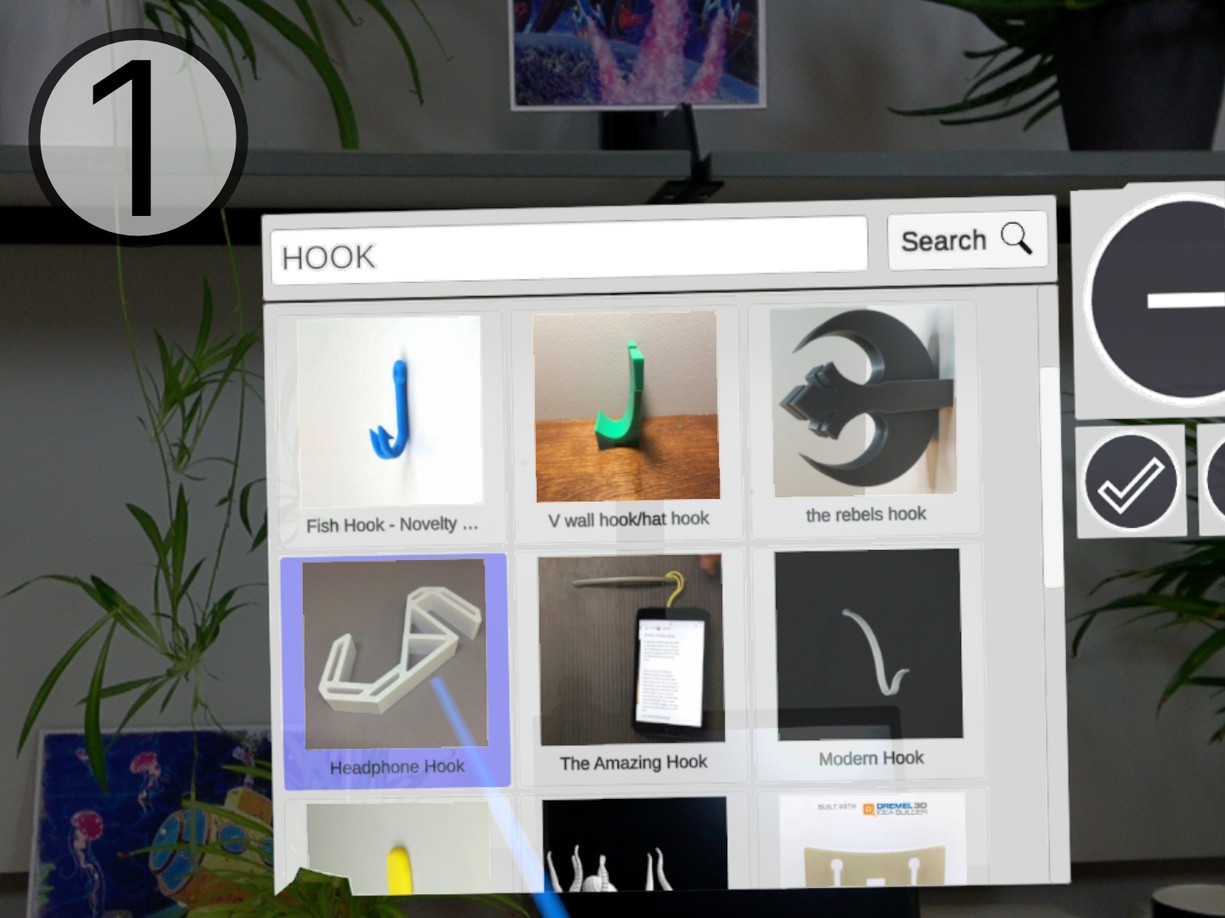}\hspace*{\fill}
                \includegraphics[width=0.328\linewidth]{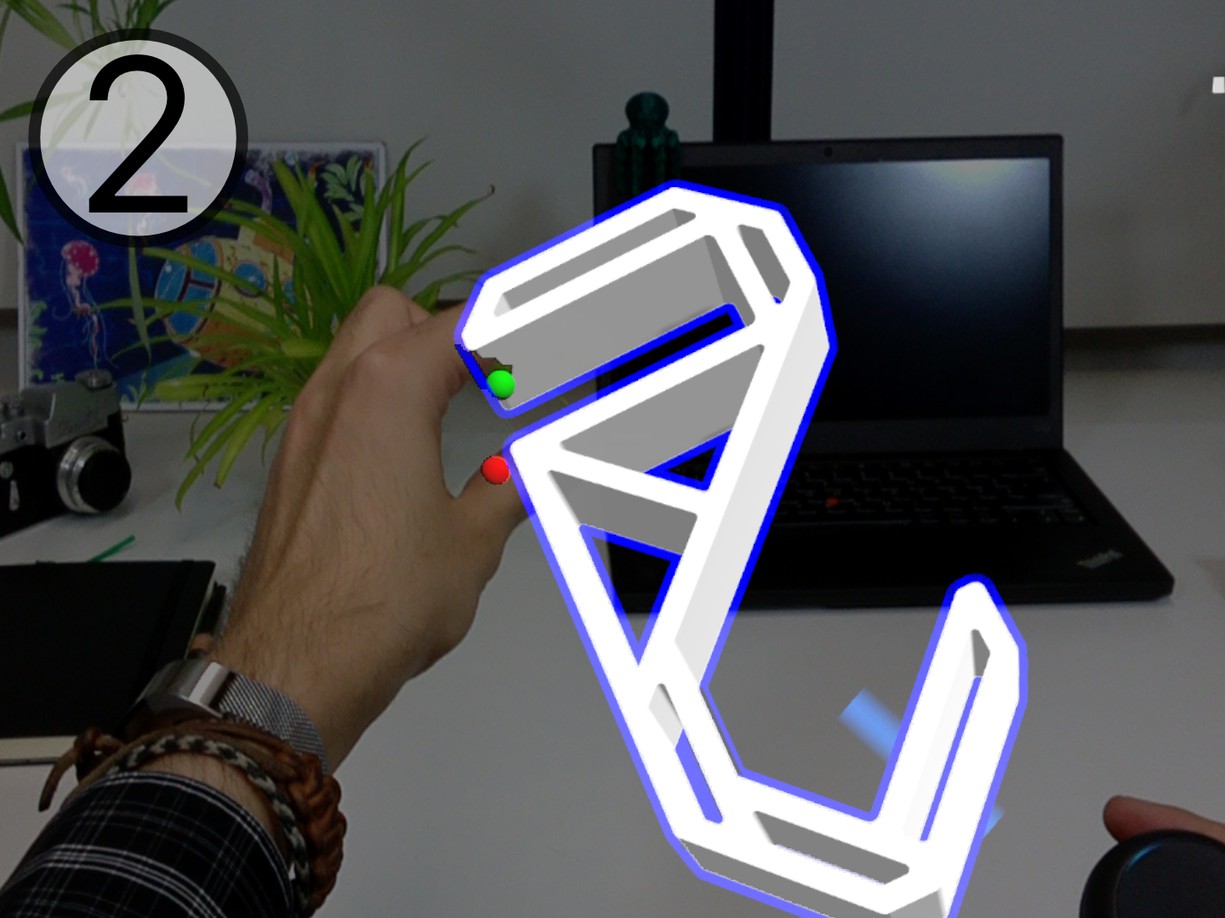}\hspace*{\fill}
                \includegraphics[width=0.328\linewidth]{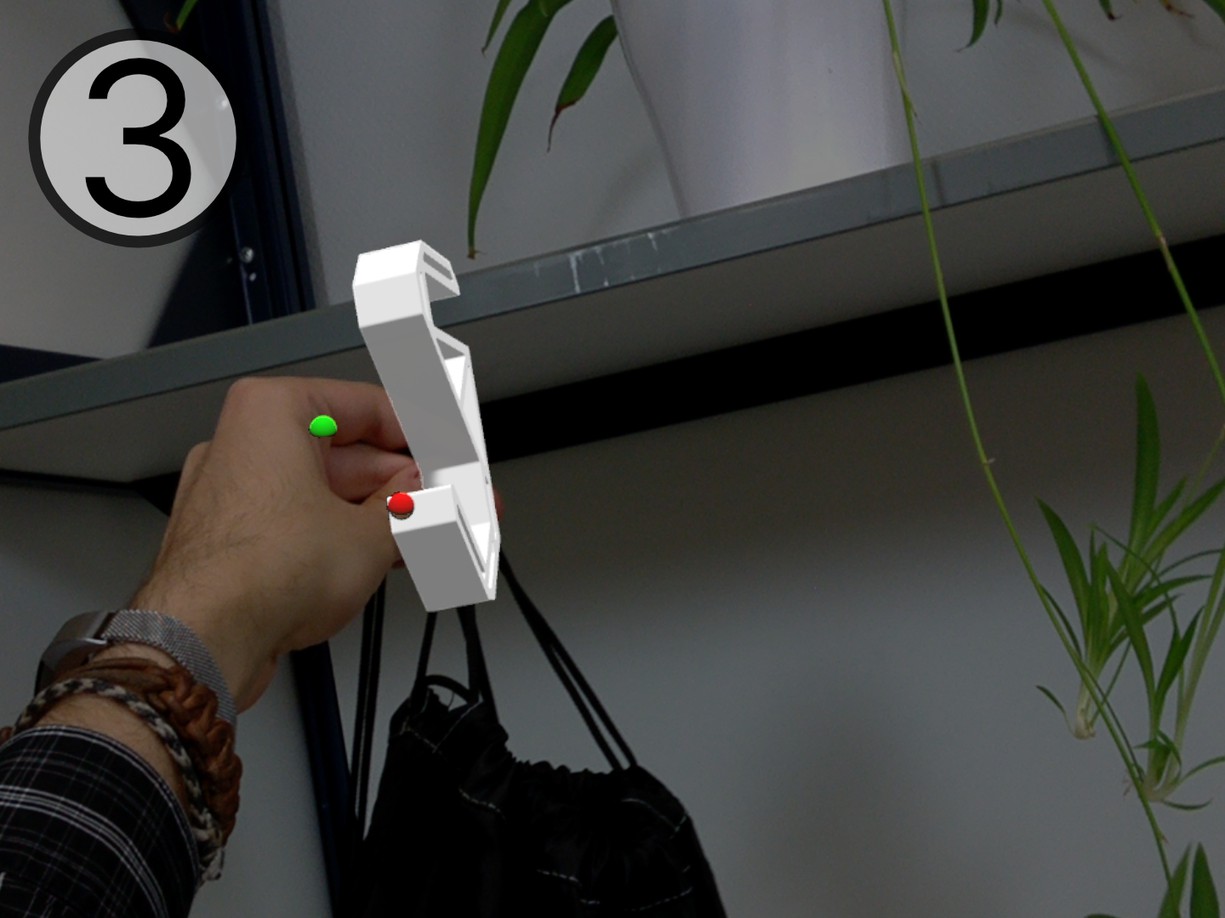}\hspace*{\fill}
                \caption{
                    Choosing and altering an existing design.
                    Browsing through hook designs (1), scaling of a fitting one (2), in-situ verification that the hook would be mounted high enough (3). 
                }
                \label{fig:hook-p-1}
            \end{figure}
            The simplest path is seen in Figure \ref{fig:hook-p-1}, where the user initiates a search for ''hook''. 
            This not only yields cloth hooks, but also hooks for headphones or wires.
            She then proceeds to select one that originally was meant for headphones, but which seems robust enough to hold a coat or a bag. 
            Finally, she takes a real bag to verify that the currently still virtual hook is high enough for the bag to hang above the table level.
            
        \subsubsection{Path 2 - Repurposing/Misusing/Remixing a Design}
            \begin{figure}[h!]
                \includegraphics[width=0.328\linewidth]{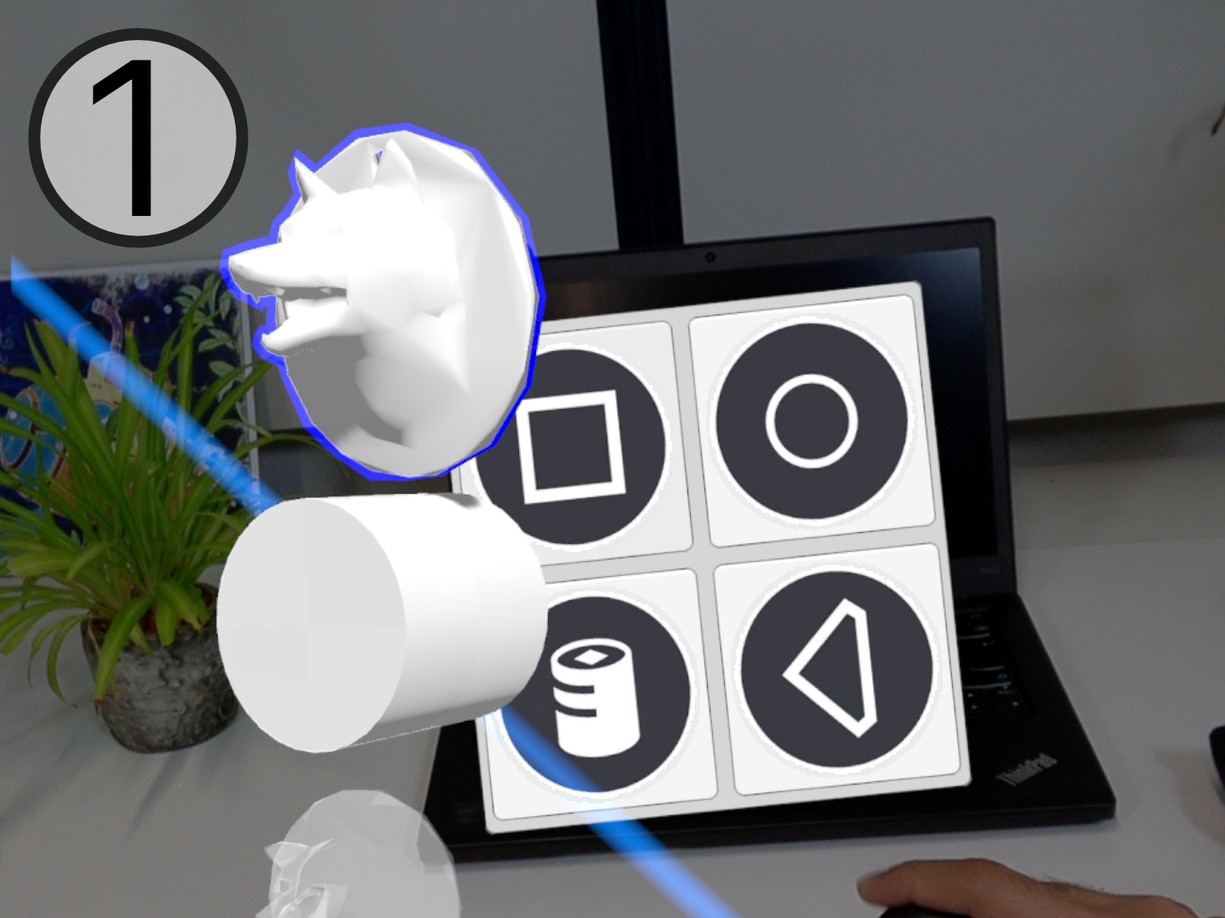}\hspace*{\fill}
                \includegraphics[width=0.328\linewidth]{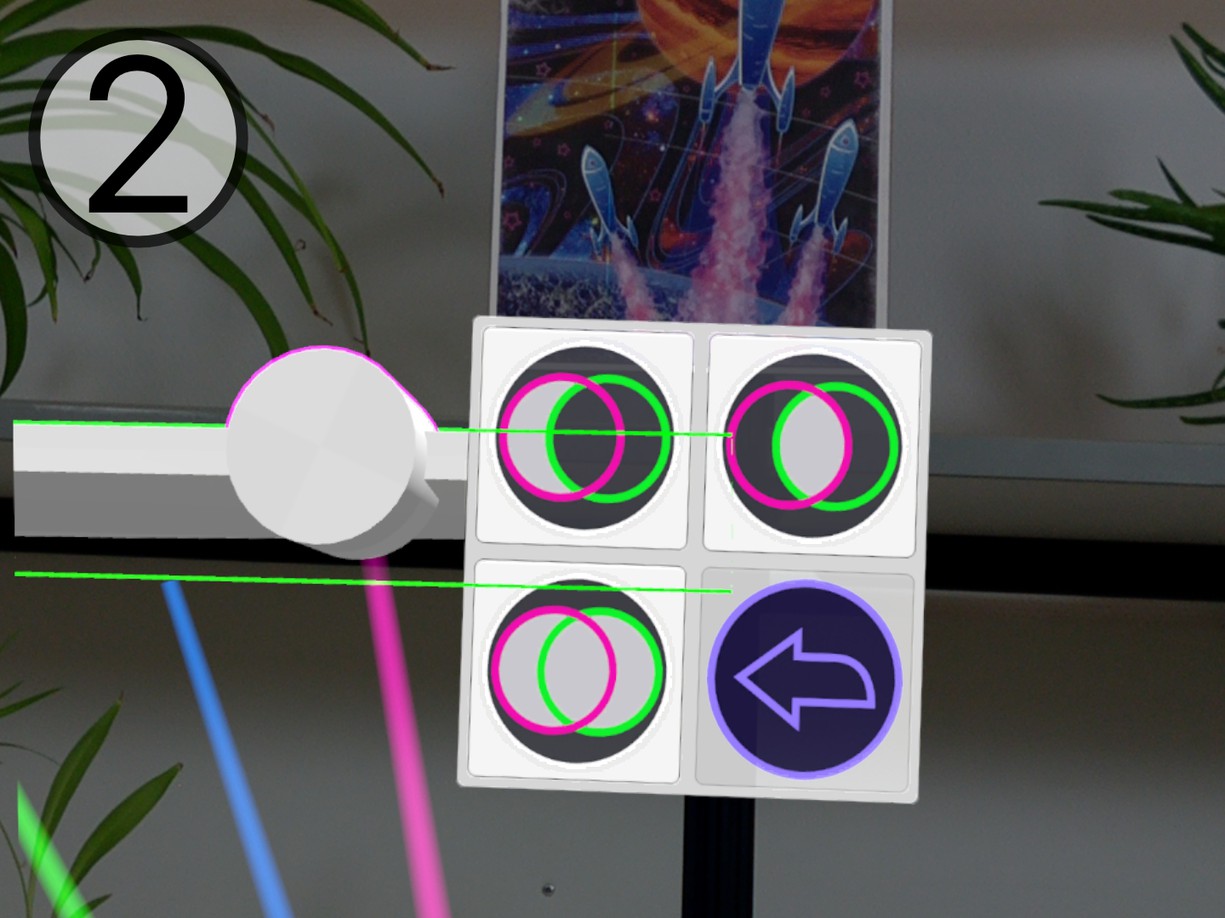}\hspace*{\fill}
                \includegraphics[width=0.328\linewidth]{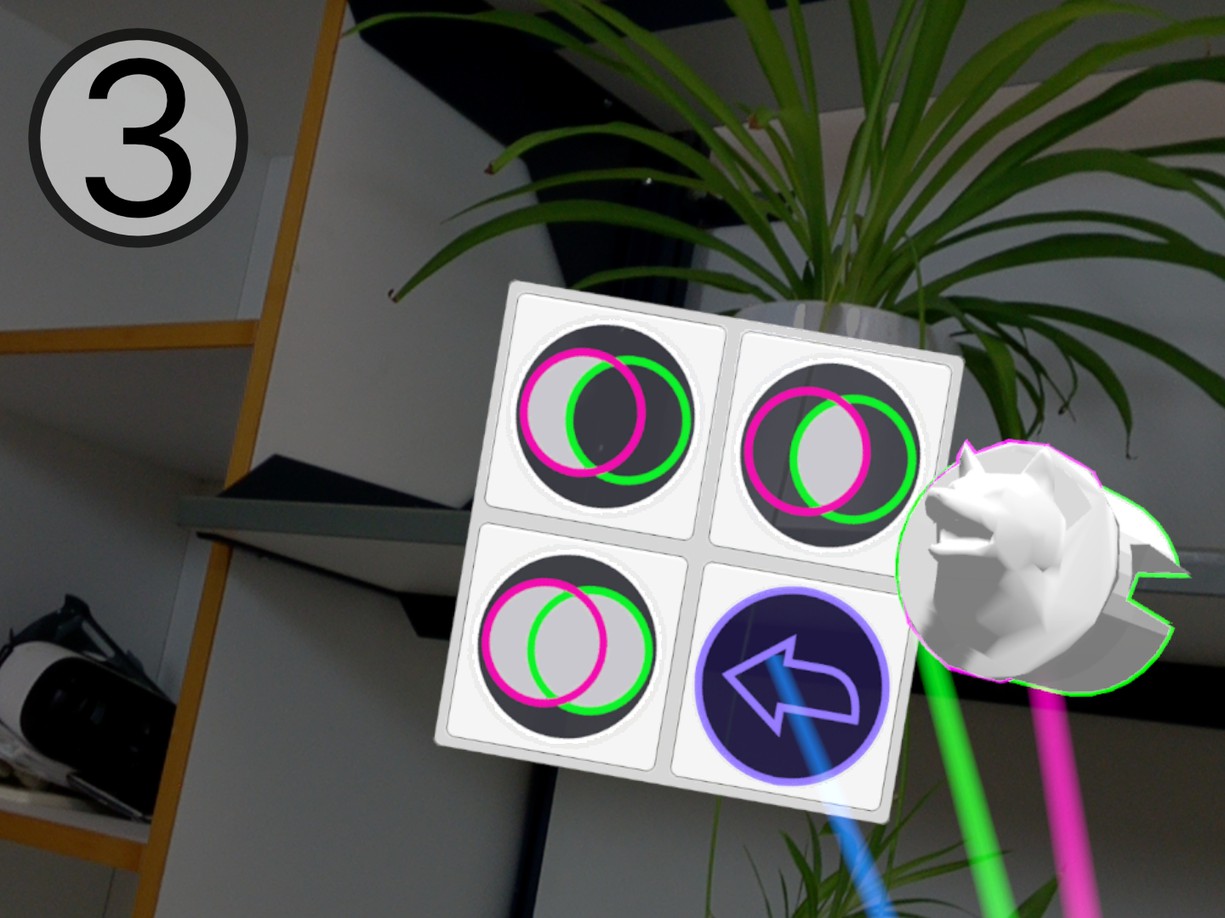}\hspace*{\fill}
                \caption{
                        Converting an animal pendant to serve as a cloth hook.
                        Cylinder primitive and the pendant as base elements (1), subtraction of the shelf from the base cylinder (2), union of the mounting cylinder and the pendant (3).
                    }
                \label{fig:hook-p-2}
            \end{figure}
            The user may likewise remix entirely different designs to achieve her goals (Figure \ref{fig:hook-p-2}).
            She sees a pendant that is meant to be worn as jewellery, depicting an animal head.
            As it is too thin to be directly mounted to the shelf, the user instantiates a primitive cylinder, as provided by \system.
            This cylinder serves as the core mounting material to the shelf.
            Afterwards, she applies the subtraction CSG operation, to cut out a portion of the shelf from the cylinder.
            Lastly, she uses the union operation to combine the mounting cylinder and the pendant for a novel coat hook. 
            As before, she can also try to verify the functionality (i.e., the height the objects will hang at) visually, prior to printing.
            
        \subsubsection{Path 3 - Replicating and Altering an Existing Artifact}
            \begin{figure}[h!]
                \includegraphics[width=0.328\linewidth]{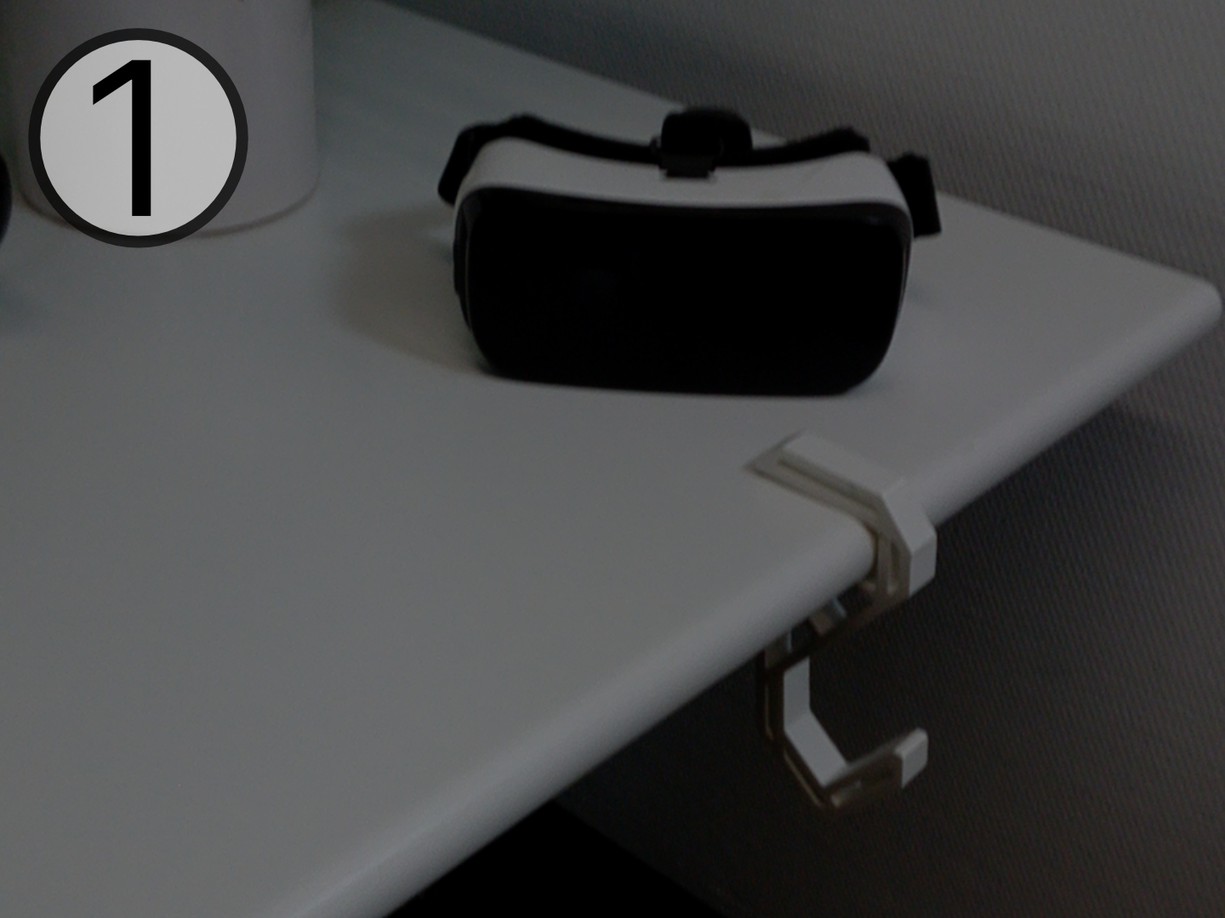}\hspace*{\fill}
                \includegraphics[width=0.328\linewidth]{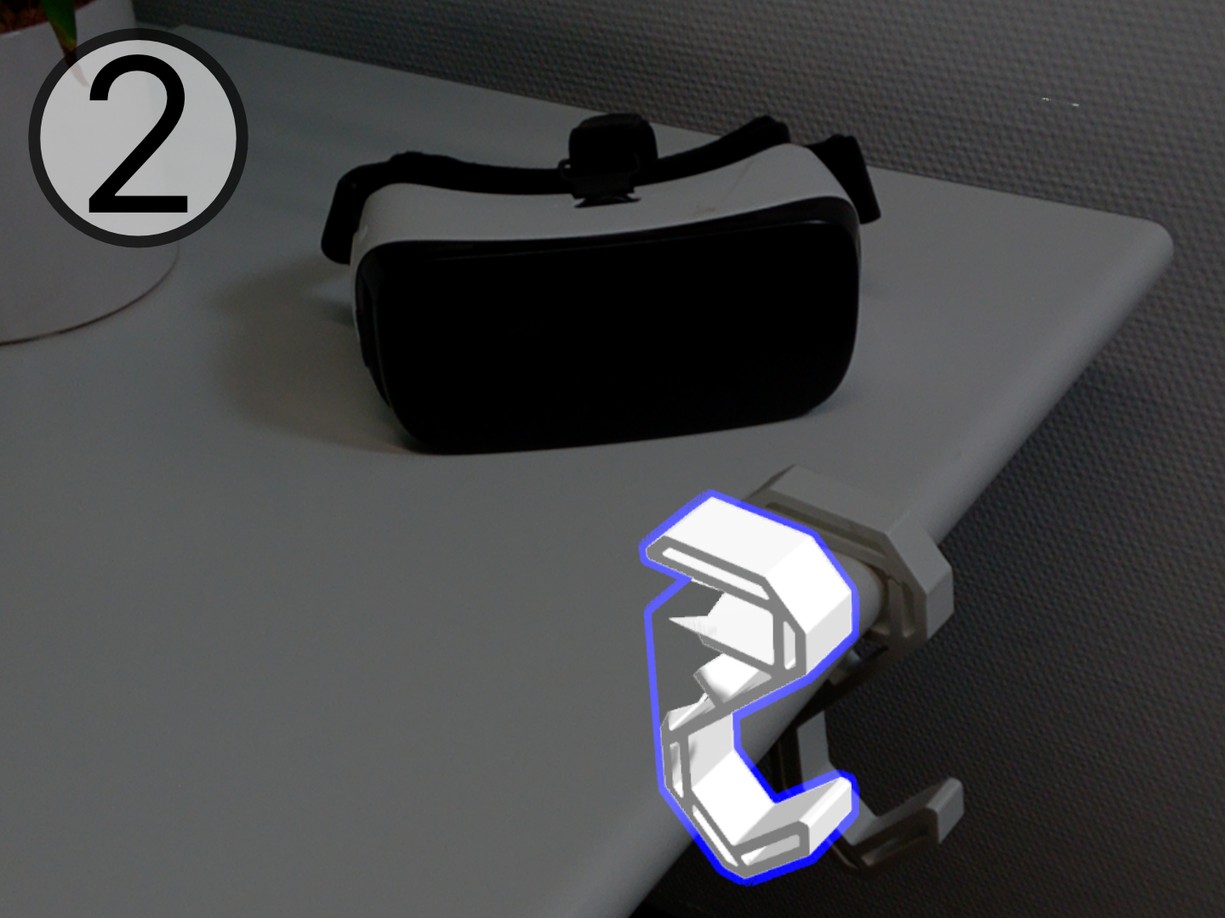}\hspace*{\fill}
                \includegraphics[width=0.328\linewidth]{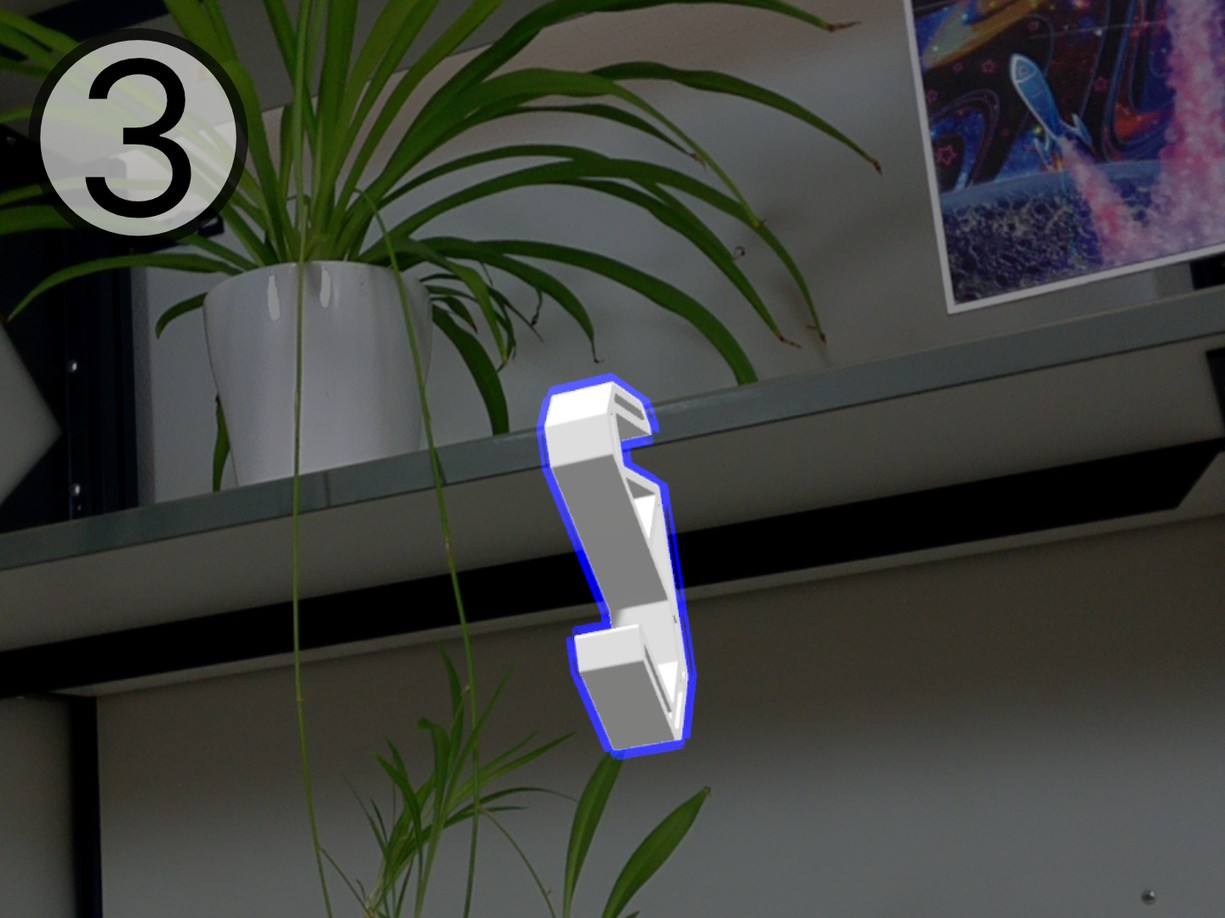}\hspace*{\fill}
                \caption{
                        Replicating an existing hook.
                        Selection of an existing hook (1), duplication and scaling of the hook (2) to achieve a fitting result (3).
                    }
                \label{fig:hook-p-3}
            \end{figure}
            Lastly, the user may leverage her own physical environment by copying and pasting an existing real-world artifact.
            This existing hook has already proven its function and provides a reference concerning a viable height for it to be mounted.
            While it may not be shelf-mounted, the user can convert it in the same fashion as the previously presented path, by either scaling it down, or alternatively adding a padding for the mount to fit the thinner shelf.

 \section{Future Work}
 \label{sec:futurework}
    Apart from conducting detailed usability evaluations with the presented system, expanding the scope of this concept to subtractive manufacturing is a conceivable next step.
    While processes like CNC milling likewise start with a 3D model, their foundational part is the material stock.
    One could either omit the concept of a material stock, or treat objects in the users physical context as stock for subtractive manufacturing.
    The alteration of artifacts, either from a model repository or the users' environment would progress in the same fashion.
    Likewise, processes that venture beyond shape remixing and instead involve more complex remixing procedures (e.g., remixing of machines~\cite{roumenGrafterRemixing3DPrinted2018}) are intriguing to consider in-situ.
    
    A reasonable extension of \systemEmph  would be the introduction of a ''snapping'' feature to support users with object alignment~\cite{nuernbergerSnapToRealityAligningAugmented2016}.
    Beyond that, any feature that supports users with aspects like scaling, orienting, coloring~\cite{jinPhotoChromeleonReProgrammableMultiColor2019a} of artifacts or with any other type of remix procedure, is a viable extension of the \paradigm-paradigm.
    Likewise, additional error tolerance could be introduced through constructs like springs~\cite{roumenSpringFitJointsMounts2019} or automated generation of connectors~\cite{koyamaAutoConnectComputationalDesign2015}.
    Coloring in particular is a relevant feature, as it depends on the available fabrication process, but heavily influences the \emph{aesthetic interaction} between the physical context and the artifact.
    \system aimed do provide an interface to the repository, but abstracted away the specifics of search.
    Filters and different ordering options were removed for clarity.
    On a more conceptual level, one could consider a more refined search feature, where specific features of artifacts (e.g., tooth counts of cogs) could be searched for.
    Physna is such a concept for a ''geometric search engine'', but targeted at industrial users~\cite{physnainc.ShapeSearchPhysna2019}.
    \system did not incorporate user-centered ways to 3D-scan objects with a HMD.
    For an ideal scan, users would have to be guided to move around the object (i.e., be the sensor), or rotate the object themselves (i.e., be the turntable).

\section{Conclusion}
\label{sec:conclusion}
    We presented \system, a tool to allow users to remix artifacts retrieved from model repositories and the physical context in-situ.
    It supports the proposed notion of the in-situ \paradigm-paradigm.
    \system bridges the disconnect between the users' physical context and the artifacts found in both digital model repositories and the users' real environment.

    Model repositories are an incredibly valuable resource for both novices and experienced makers.
    By delegating the design effort of various artifacts, the maker may focus on aspects of customization and personalization of these artifacts -- refining, remixing and tailoring them.
    As such, few problems that have never been solved before will be met by makers.
    However, the intricate specifics of functional and aesthetic fit are often unique enough to warrant either the adaptation of existing artifacts or the design of entirely new ones.
    Both paths require the investment of time for both novices and experienced users: learning tools, measuring the environment, adapting or creating designs.
    \system, is a mixed-reality-based tool to allow users to alter and remix artifacts retrieved from model repositories in-situ.
    \system not only utilizes the remote, digital repository as a source for artifacts and features, but also the users' physical context.
    This bridges the disconnect between the users' unique physical context, and the versatile offers model repositories can make, making it easier to omit the process of modelling, while retaining predictable and appropriate results.

\section{Acknowledgements}
We thank Ali Askari and Jan Rixen for their feedback and thoughtful discussions.

\balance{}

\bibliographystyle{SIGCHI-Reference-Format}
\bibliography{zotero}

\end{document}